\documentclass[11pt]{article}
\usepackage{fullpage}
\usepackage[latin1]{inputenc}
\usepackage{graphicx}
\usepackage{epsfig}
\usepackage{color}
\usepackage{enumerate}
\usepackage{verbatim}
\usepackage{amsthm, amssymb}
\usepackage{amsmath}
\usepackage{mdwlist}

\usepackage[
normalsections, 
normalmargins, 
normalfloats, 
normalindent, 
normaltitle, 
normalleading, 
normalbib, 
normalbibnotes, %
]{savetrees}

\newtheorem{theorem}{Theorem}
\newtheorem{lemma}[theorem]{Lemma}

\newtheorem{corollary}[theorem]{Corollary}

\newcommand{\intc}{\ensuremath{\mathit{int}}}
\newcommand{\extc}{\ensuremath{\mathit{ext}}}

\def\polylog{\operatorname{polylog}}

\def\dist{\delta}

\newenvironment{mylist}[1]{
\setbox1=\hbox{#1}
\begin{list}{}{
\topsep=0.02in\itemsep=0in\parsep=0.02in\partopsep=0.1in
\setlength{\labelwidth}{\wd1}
\setlength{\leftmargin}{\wd1}
\addtolength{\leftmargin}{0.5em}
\addtolength{\leftmargin}{\labelsep}
\setlength{\rightmargin}{0.5em}
}}{\end{list}}

\newcommand{\litem}[1]{\item[#1\hfill]}

\newcommand{\out}[1]{}

\title{\bf Improved Minimum Cuts and Maximum Flows in Undirected Planar Graphs}
\author{
Giuseppe F. Italiano\thanks{Dipartimento di Informatica, Sistemi e Produzione, University of Rome "Tor Vergata", {\tt italiano@disp.uniroma2.it}. Work partially supported by
the 7th Framework Programme of the EU  (Network of Excellence ``EuroNF: Anticipating the Network of the Future - From Theory to Design'')
and by MIUR,
the Italian Ministry of Education, University and Research, under Project
AlgoDEEP.}
\and
Piotr Sankowski\thanks{Institute of Informatics,
University of Warsaw and Department of Computer and System Science,
Sapienza University of Rome, {\tt sank@mimuw.edu.pl}.}
}

\begin{document}

\maketitle

\thispagestyle{empty}
\begin{abstract}
In this paper we study minimum cut and maximum flow problems on planar graphs, both in static and in dynamic settings. First, we present an algorithm that
given an undirected planar graph computes the minimum cut between any two given vertices in $O(n \log \log n)$ time.
Second, we show how to achieve the same $O(n \log \log n)$ bound for the problem of computing maximum flows in undirected planar graphs. To the best of our knowledge, these are the first algorithms
 for those two problems that break the $O(n \log n)$ barrier,  which has been standing for more than 25 years.
Third, we present a fully dynamic algorithm that is able to maintain information about minimum cuts and maximum flows in a plane graph (i.e., a planar graph with a fixed embedding): our algorithm is able to insert edges, delete edges and  answer min-cut and max-flow queries between any pair of vertices in $O(n^{2/3} \log^3 n)$\out{\footnote{Throughout the paper, we use $\widetilde{O}(f(n))$ to denote $O(f(n)\polylog(n))$.}} time per operation. This result is based on a new dynamic shortest path algorithm for planar graphs which may be of independent interest. We remark that this is the first known non-trivial algorithm for min-cut and max-flow problems in a dynamic setting.
\end{abstract}

\newpage

\addtocounter{page}{-1}

\section{Introduction}
\label{section:introduction}

Minimum cut and maximum flow problems have
been at the heart of algorithmic research on graphs for over 50 years.
Particular attention
has been given to solving those problems
on planar graphs, not only because
they
often admit faster algorithms than general
graphs but also since planar graphs arise naturally in
many applications.
The pioneering work of Ford and Fulkerson~\cite{ff56,ff62},
which introduced the max-flow min-cut theorem,
also contained an elegant
algorithm for computing maximum flows in {\em $(s,t)$-planar} graphs (i.e., planar graphs where both the source $s$   and the sink $t$ lie on the same face). The algorithm was implemented to work in $O(n \log n)$ time by Itai and
Shiloach~\cite{itai-shiloach-79}. Later, a simpler algorithm for the same problem was given by Hassin~\cite{hassin-81},
who reduced the problem to single-source shortest path computations in the dual graph. The time required to compute single-source shortest paths in
$(s,t)$-planar graphs
was shown to be $O(n\sqrt{\log n}\,)$ by Frederickson~\cite{federickson-87} and later improved to $O(n)$
by Henzinger {\it et al.}~\cite{henzinger-et-al-97}. As a result, minimum cuts and maximum flows can be found in $O(n)$ time in $(s,t)$-planar graphs.

Itai and Shiloach~\cite{itai-shiloach-79} generalized their approach to the case
of general planar (i.e., not only $(s,t)$-planar) graphs, by observing that
the minimum cut separating vertices $s$ and $t$ in a planar graph $G$ is related to the minimum cost cycle that separates faces $f_s$ and $f_t$ (corresponding to vertices $s$ and $t$) in the dual graph.
The resulting
algorithm
makes $O(n)$ calls to their original algorithm for $(s,t)$-planar graphs
and thus runs in a total of $O(n^2 \log n)$ time.
In the case of undirected planar graphs,
Reif~\cite{Reif83}
improved this
bound by describing how to find the minimum cost separating cycle with a divide-and-conquer approach
using only $O(\log n)$ runs of the $(s,t)$-planar
algorithm: this yields an
 $O(n \log^2 n)$ time to compute a minimum cut for undirected planar graphs.
Later on, Frederickson~\cite{federickson-87} improved the
running time of Reif's algorithm to $O(n \log n)$. The same result can be obtained by using more recent planar shortest path algorithms (see e.g.,~\cite{henzinger-et-al-97}).
Hassin and Johnson~\cite{hassin-johnson-85} extended the minimum cut algorithm of Reif to
compute a maximum flow in only $O(n \log n)$ additional
time: this implies an undirected planar
maximum flow algorithm that runs in $O(n \log n)$ time as well.
In summary, the best bound known for computing
minimum cuts and maximum flows in planar undirected graphs is $O(n\log n)$.

The first contribution of this paper is to improve to $O(n \log \log n)$ the time for computing minimum cuts in planar undirected graphs. To achieve this bound, we improve Reif's classical approach~\cite{Reif83} with several novel ideas. To compute a minimum $s$-$t$ cut in a planar graph $G$, we first identify a path $\pi$ between face $f_s$ (corresponding to vertex $s$) and face $f_t$ (corresponding to vertex $t$) in the dual graph $G_D$. Next, we compute a new graph $G_{\pi}$ as follows. We cut $G_D$ along path $\pi$, so that the vertices and edges of $\pi$ are duplicated and lie on the boundary of a new face: another copy of the same cut graph is embedded inside this face. We show that minimum separating cycles in $G_D$ correspond to some kind of shortest paths in $G_{\pi}$.  Applying a divide-and-conquer approach on the path $\pi$ yields the same $O(n\log n)$ time bound as previously known algorithms~\cite{federickson-87,hassin-johnson-85,Reif83}. However, our novel approach has the main advantage that it allows the use of {\it any} path $\pi$ in the dual graph $G_D$, while previous algorithms  were constrained to choose $\pi$ as a shortest path. We will exploit the freedom implicit in the choice of this path $\pi$ to produce a faster $O(n\log\log n)$ time algorithm, by using a suitably defined cluster decomposition of a planar graph, combined with the Dijkstra-like shortest path algorithm by Fakcharoenphol and Rao~\cite{fr06}.
Our second contribution is to show that also maximum flows can be computed in undirected planar graphs within the same $O(n \log \log n)$ time bound. We remark that this is not an immediate consequence of our new minimum cut algorithm: indeed the approach of Hassin and Johnson~\cite{hassin-johnson-85} to extend minimum cut algorithms  to the problem of computing maximum flows has a higher overhead of $O(n\log n)$. To get improved maximum flow algorithms,
we have to appropriately modify the original technique of Hassin and Johnson~\cite{hassin-johnson-85}.
To the best of our knowledge,
the algorithms presented in this paper are the first algorithms
that break the $O(n \log n)$ long-standing barrier for minimum cut and maximum flow problems in undirected planar graphs.

As our third contibution, we present a fully dynamic algorithm that is able to maintain information about minimum cuts and maximum flows in a plane graph (i.e., a planar graph with a fixed embedding): our algorithm is able to insert edges, delete edges and  answer min-cut and max-flow queries between any pair of vertices in $O(n^{2/3}\log^3 n)$ time per operation. This result is based on
the techniques developed in this paper for the static minimum cut algorithm
and on a new dynamic shortest path algorithm for planar graphs which may be of independent interest. We remark that this is the first known non-trivial algorithm for min-cut and max-flow problems in a dynamic setting.

\section{Minimum Cuts in Planar Graphs}
\begin{figure}[htp]
\centerline{\psfig{file=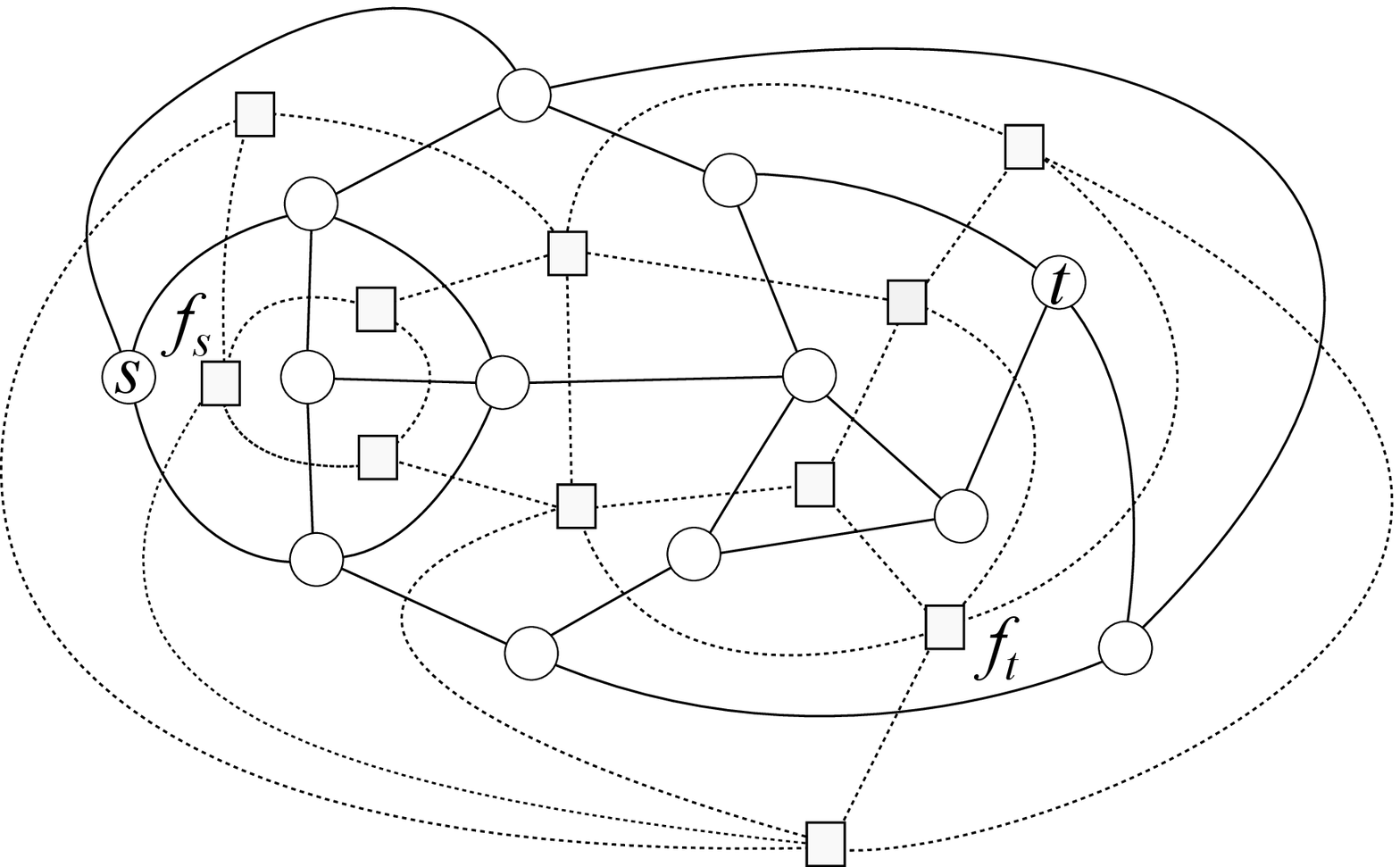,width=12cm}}
\caption{An embedded planar graph $G$ and its dual graph $G_D$. Vertices of $G$ are shown as circles, and edges of $G$ are solid. Vertices of $G_D$ are shown as gray squares, and edges of $G_D$ are dashed. $s$ and $t$ are two vertices in $G$, and $f_s$ and $f_t$ are arbitrary inner faces incident respectively to $s$ and $t$.}
\label{fig:primal-dual}
\end{figure}

Let $G=(V,E,c)$ be a planar undirected graph where $V$ is the vertex set, $E$ is the edge set and
$c:E \to \mathcal{R}^+$ is the edge capacity function.
Let the planar graph $G$ be given with a certain embedding.
Using the topological incidence
 relationship between edges and
 faces of $G$, one can define the
 {\em dual graph}
 $G_D = (F,E_D,c_D )$ as follows.
Each face of $G$ gives rise to a vertex in $F$.
 Dual vertices $f_1$ and $f_2$
 are connected by a dual undirected edge $e_D $
 whenever primal edge $e$
 is adjacent to the faces of $G$
 corresponding to $f_1$
 and $f_2$. The weight $c_D(e_D)$ of the dual edge $e_D$ is equal to the weight $c(e)$ of the primal edge:  $c_D(e_D)$ is referred to as the length of edge $e_D$. In other terms,
the length of the dual edge $e_D$ is equal to the capacity of the primal edge $e$.
 In the following, we refer to $G$ as the {\it primal} graph and to $G_D$ as its {\it dual}. Throughout the paper we will refer to vertices of the dual graph $G_D$ interchangeably as (dual) vertices or faces.
 Note that $G_D$
 can be embedded in the plane by
 placing each dual vertex inside
 the corresponding face of $G$,
 and placing dual edges so that
 each one crosses only its corresponding primal edge. Thus,
 the dual graph is planar as well, although it might contain multiple edges and self loops.
  In simpler terms, the dual graph $G_D$ of a planar embedded graph $G$
 is obtained by exchanging the roles of faces and vertices,
 and $G$ and $G_D$ are each other's dual. Figure~\ref{fig:primal-dual} shows an embedded planar graph $G$ and its dual $G_D$.


Let $s$ and $t$ be any two vertices of $G$ (not necessarily on the same face). We consider the problem of finding a minimum cut in $G$ between vertices $s$ and $t$. Let  $C$ be a cycle of graph $G$:
we define the {\it interior} of $C$, denoted by $\intc(C)$, to be the region inside $C$ and including $C$ in the planar embedding of the graph $G$. We can define  the {\it exterior} $\extc(C)$ of the cycle $C$ in a similar fashion. A cycle $C$ in $G_D$ is said to be a {\it cut-cycle} if $\intc(C)$ contains exactly $s$ but not $t$. The following lemma was proven by Johnson~\cite{johnson-87}.

\begin{lemma}
\label{lem:Johnson}
A minimum $s$-$t$ cut in $G$ has the same cost as a minimum cost cut-cycle of $G_D$.
\end{lemma}

The lemma follows by the observation that for any cut-cycle $C$ the faces of $G_D$ inside $\intc(C)$ give a set of vertices $S$ in $G$ which defines a cut separating $s$ and $t$.
Note that Lemma~\ref{lem:Johnson} gives an equivalence between min-cuts in the primal graph $G$ and minimum cost cut-cycles in the dual graph $G_D$.
By using a divide-and-conquer approach, this equivalence
can be turned into an efficient algorithm for finding flows in undirected planar graphs~\cite{Reif83}. The resulting algorithm, combined with more recent results on shortest paths in planar graphs~\cite{henzinger-et-al-97}, is able to work in a total of $O(n \log n)$ time. However, this approach seems to inherently require $O(n \log n)$ time, and does not seem to leave margin for improvements. In the next section, we will present a completely different
and more flexible approach, which will yield faster running times.


\subsection{Computing Min-Cuts}
\label{section-computing-min-cuts}

\begin{figure}[thp]
\centerline{
\begin{tabular}{c}
\psfig{file=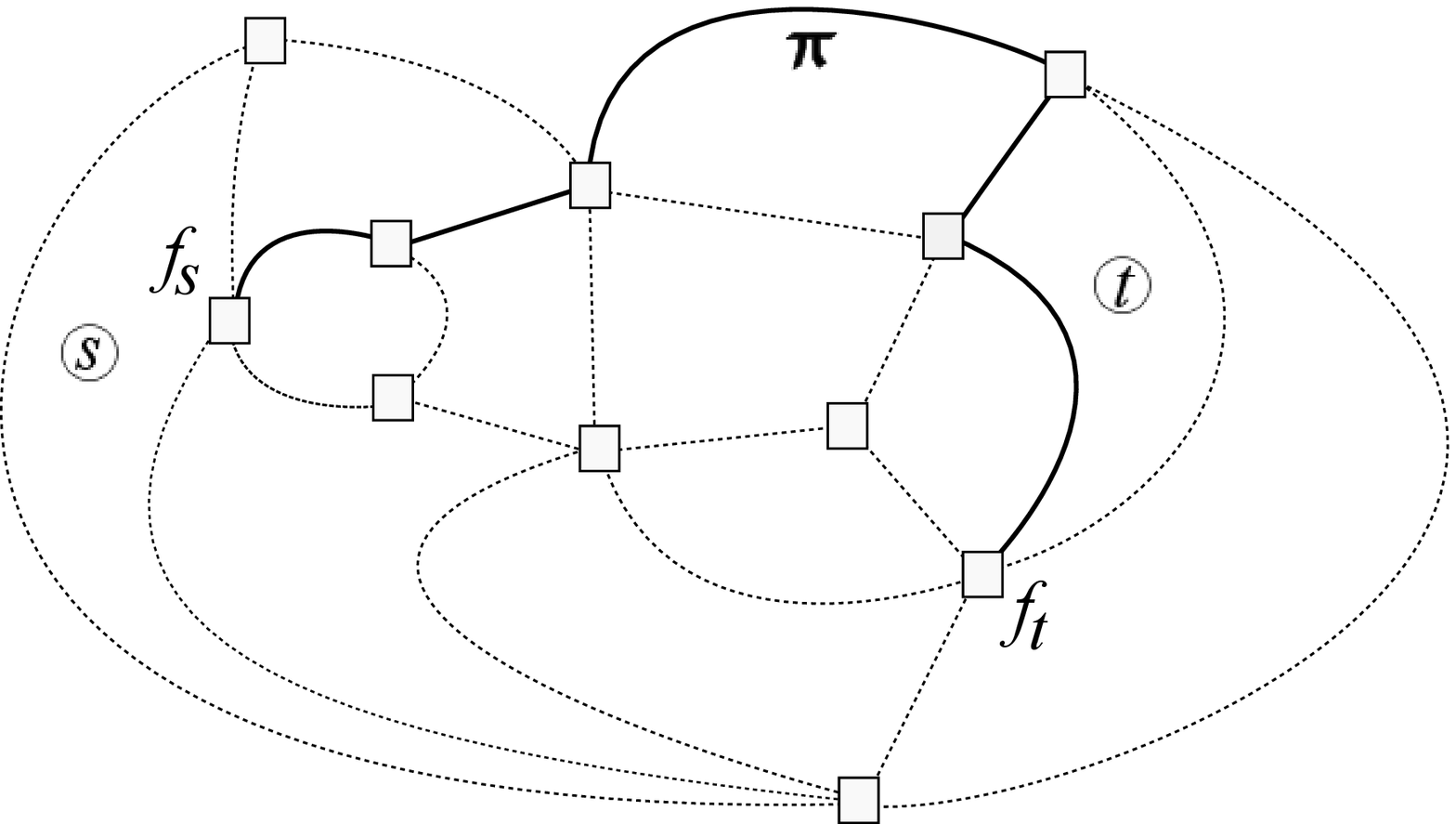,width=7cm}\\(a)\\
\end{tabular}
\hspace{7mm}
\begin{tabular}{c}
\psfig{file=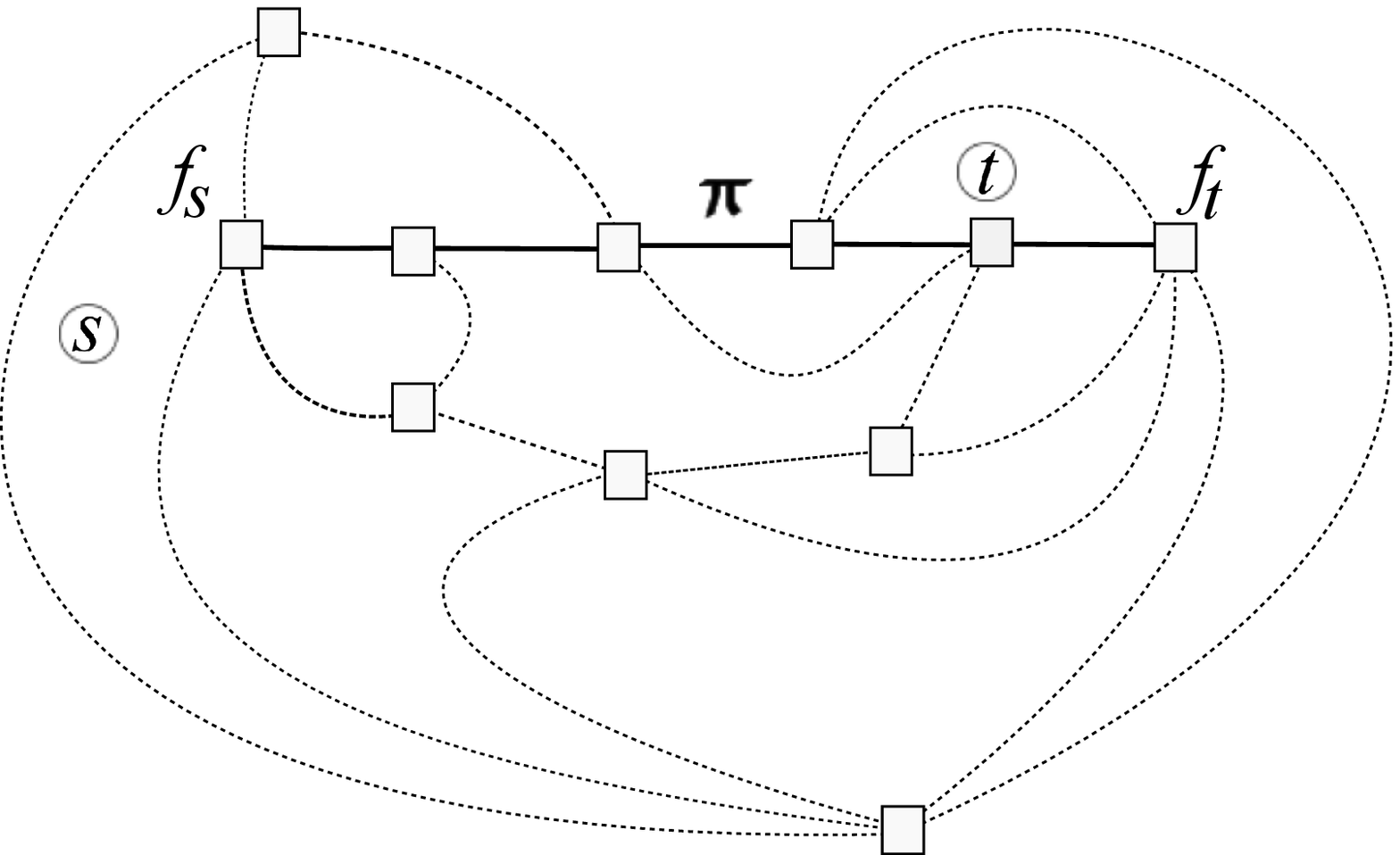,width=7cm}\\(b)\\
\end{tabular}
}
\vspace{5mm}
\centerline{
\begin{tabular}{c}
\psfig{file=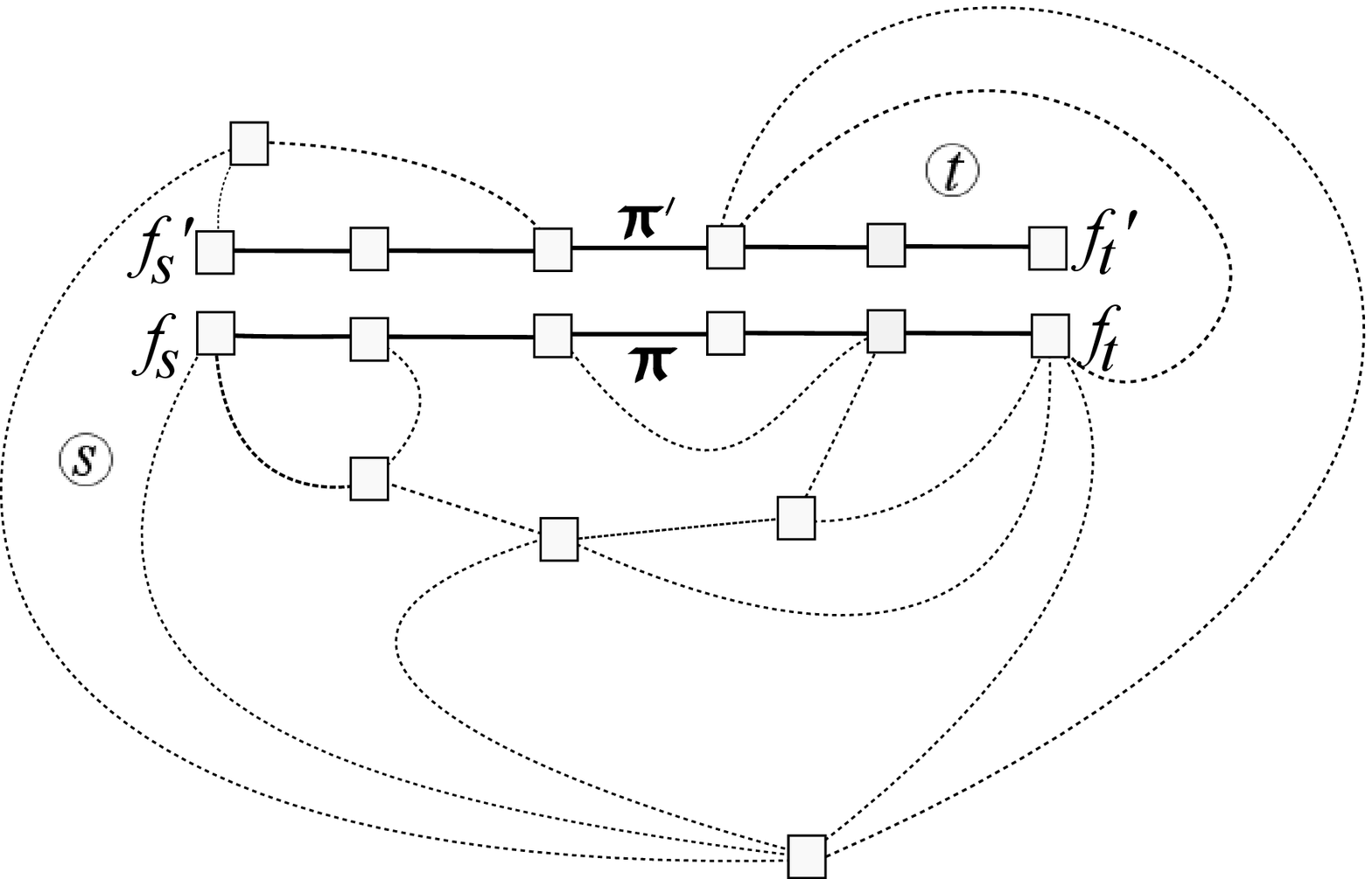,width=7cm}\\(c)\\
\end{tabular}
\hspace{7mm}
\begin{tabular}{c}
\psfig{file=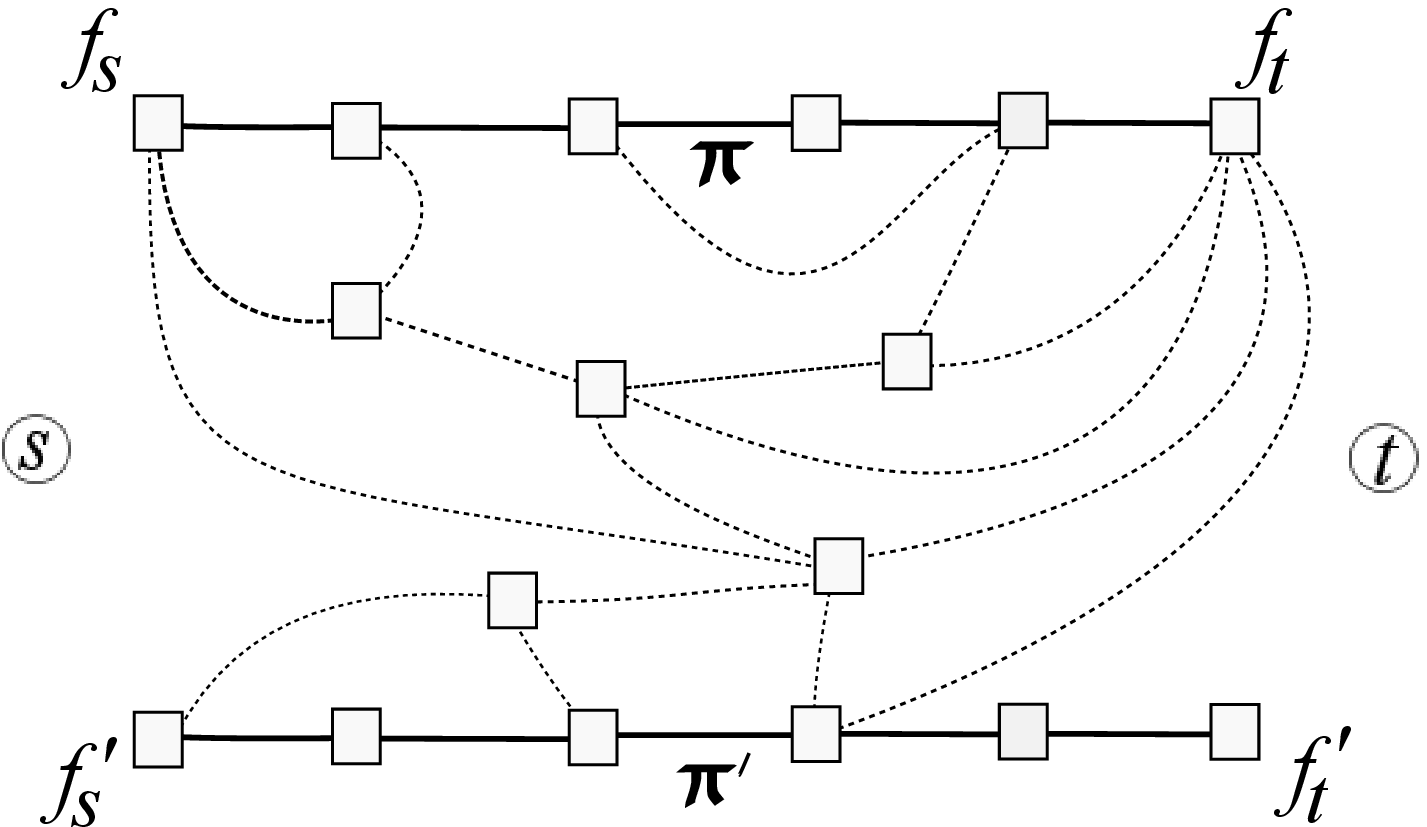,width=7cm}\\(d)\\
\end{tabular}
}
\vspace{5mm}
\centerline{\psfig{file=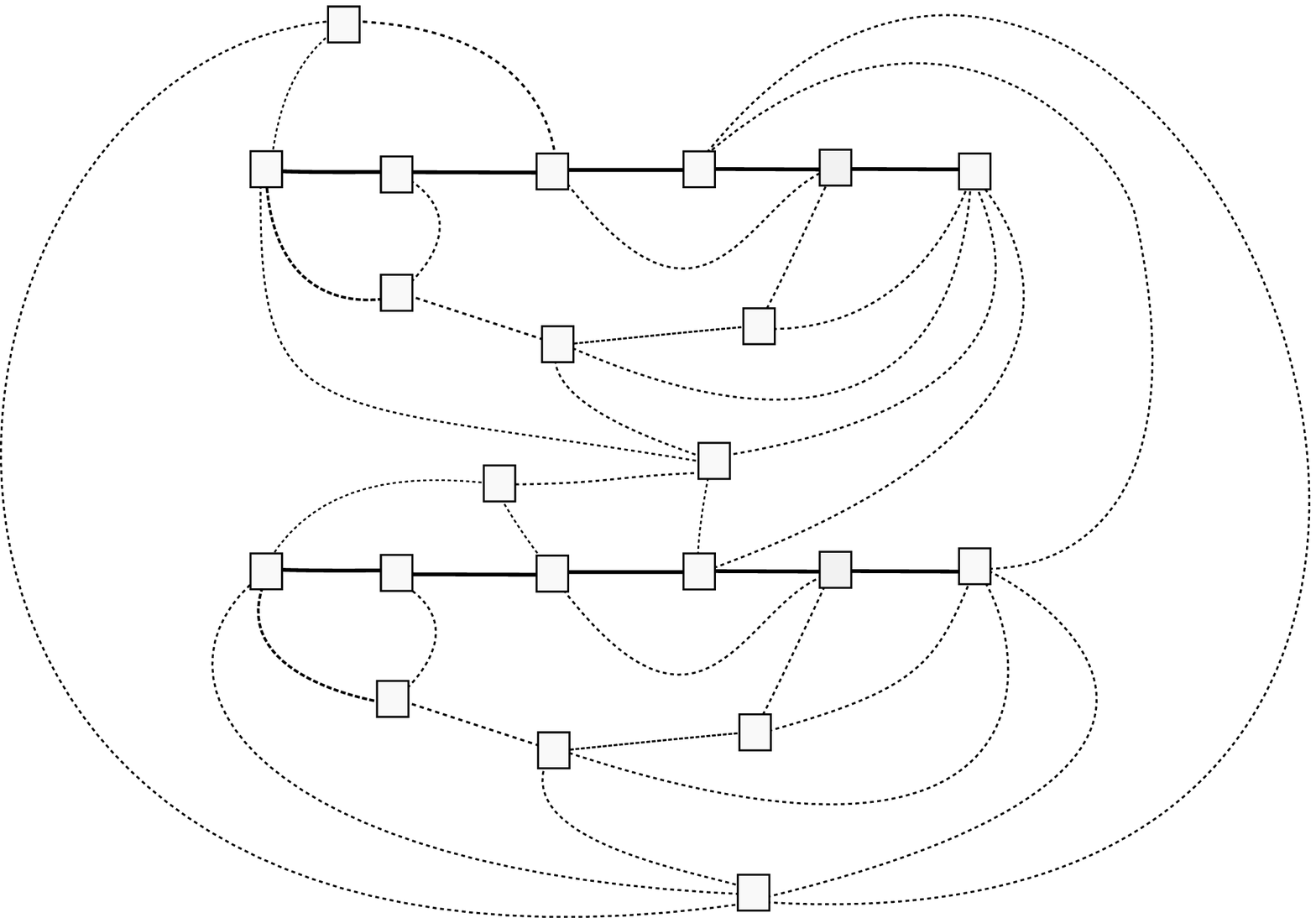,width=12cm}}
\centerline{(e)}
\caption{
(a) The dual graph $G_D$ of Figure~\ref{fig:primal-dual}: the path $\pi$ from $f_s$ to $f_t$ is shown in bold. 
(b) The dual graph $G_D$ embedded so that the path $\pi$ is a horizontal line. 
(c) The graph $G_{\pi}'$. 
(d) The graph $G_{\pi}''=G_{\pi}'$, embedded so that the face defined by $\pi$ and $\pi'$ is the outer face. 
(e) The graph $G_{\pi}$ obtained after identifying the path $\pi'$ (respectively $\pi$) in $G_{\pi}''$ with the path $\pi$ (respectively $\pi'$) in $G_{\pi}'$.}
\label{fig:Gpi}
\end{figure}

Let $f_s$ and $f_t$ be arbitrary inner faces incident to $s$ and to $t$ respectively. Find any simple path $\pi$ from $f_s$ to $f_t$ in $G_D$. The path $\pi$ can be viewed as connecting special vertices in the dual graph corresponding to $s$ and $t$. Hence, any $s$-$t$ cut needs to cross this path, because it splits $s$ from $t$. Let $\pi$ traverse dual vertices $f_1,\ldots, f_k$, where $f_1=f_s$ and $f_k=f_t$.
Let us look at the path $\pi$ as a horizontal line, with $f_s$ on the left and $f_t$ on the right (see Figure~\ref{fig:Gpi}(b)). An edge $e_D\not\in\pi$ in $G_D$ such that $e_D$ is incident to some face $f_i$, $1\leq i\leq k$, can be viewed as connected to $f_i$ from below or from above.
We now define a new graph $G_{\pi}'$,
by cutting $G_D$ along path $\pi$,
so that the vertices and edges of $\pi$ are duplicated and lie on the boundary of a new face. This is done as follows.
Let $\pi'$ be a copy of $\pi$, traversing new vertices $f_1'$, $f_2', \ldots, f_{k}'$. Then $G_{\pi}'$ is the graph obtained from $G_D$ by reconnecting to $f_i'$ edges entering $f_i$ from above, $1\leq i\leq k$ (see Figure~\ref{fig:Gpi}(c)). Let $G_{\pi}''$ be a copy of $G_{\pi}'$. Turn over the graph $G_{\pi}''$ to make the face defined by $\pi$ and $\pi'$ to be the outer face (see Figure~\ref{fig:Gpi}(d)). Now, identify vertices
on the path $\pi'$ (respectively $\pi$) in $G_{\pi}''$ with the vertices on the path $\pi$ (respectively $\pi'$) in $G_{\pi}'$. Denote the resulting graph as $G_{\pi} $ (see Figure~\ref{fig:Gpi}(e)). Note that the obtained graph $G_{\pi}$ is planar.

We define an {\it $f_i$-cut-cycle} to be a  cut-cycle in $G_D$ that includes face $f_i$ and does not include any face $f_j$ for $j>i$. The proof of the following lemma is immediate.

\begin{lemma}
\label{lem:minimum-cut-cycle}
Let $C_i$ be a minimum $f_i$-cut-cycle in $G_D$ for $i = 1,\ldots,k$. Then $C_i$ with minimum
cost is a minimum cut-cycle in $G_D$.
\end{lemma}

A path $\rho$ between $f_i$ and $f_i'$  in $G_{\pi}$ is said to be {\it $f_i$-separating} if $\rho$ contains neither $f_j$ nor $f_j'$, for $j>i$.
We say that a cycle $C$ in $G_D$ {\it touches} path $\pi$ in face $f_i$ if two edges on $C$ incident to $f_i$ go both up or both down, whereas we say that $C$ {\it crosses} $\pi$ in face $f_i$ if one of these edges goes up, whereas the other goes down.

\begin{figure}[htp]
\centerline{
\begin{tabular}{c}
\psfig{file=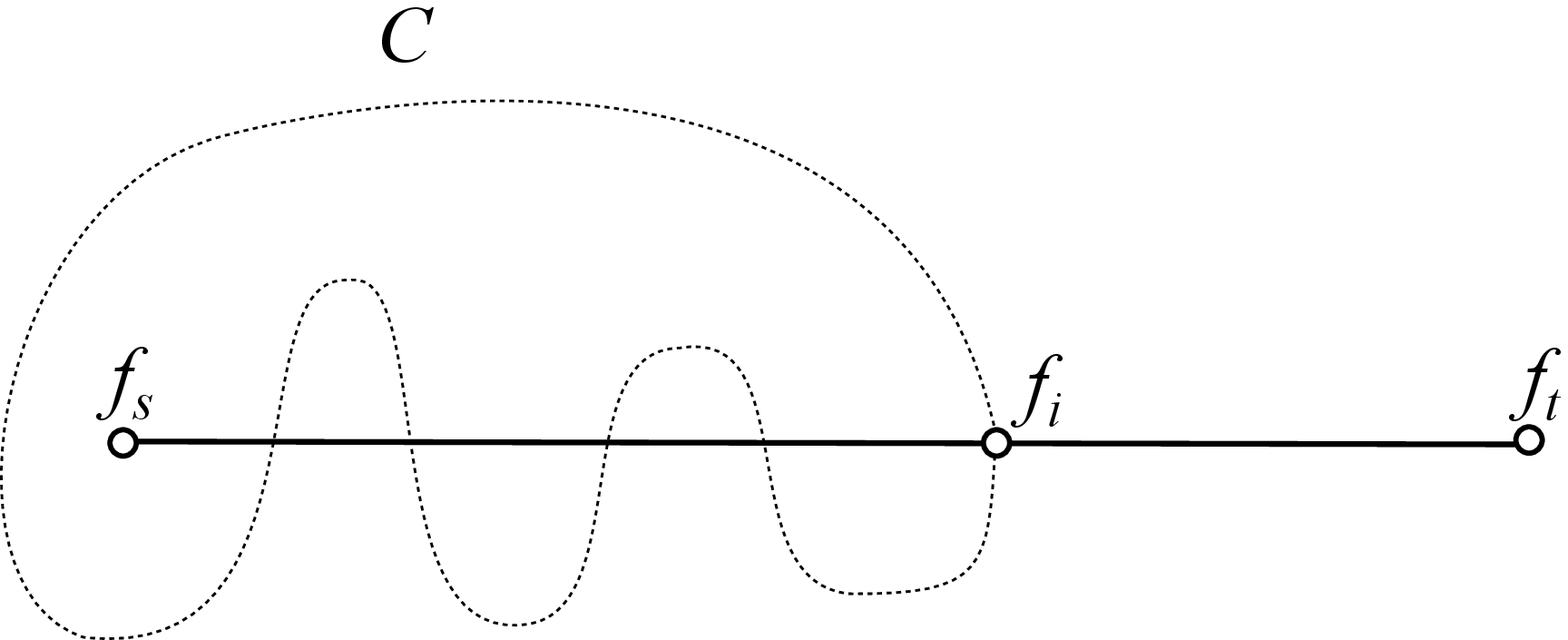,width=7cm}\\(a)\\
\end{tabular}
\hspace{7mm}
\begin{tabular}{c}
\psfig{file=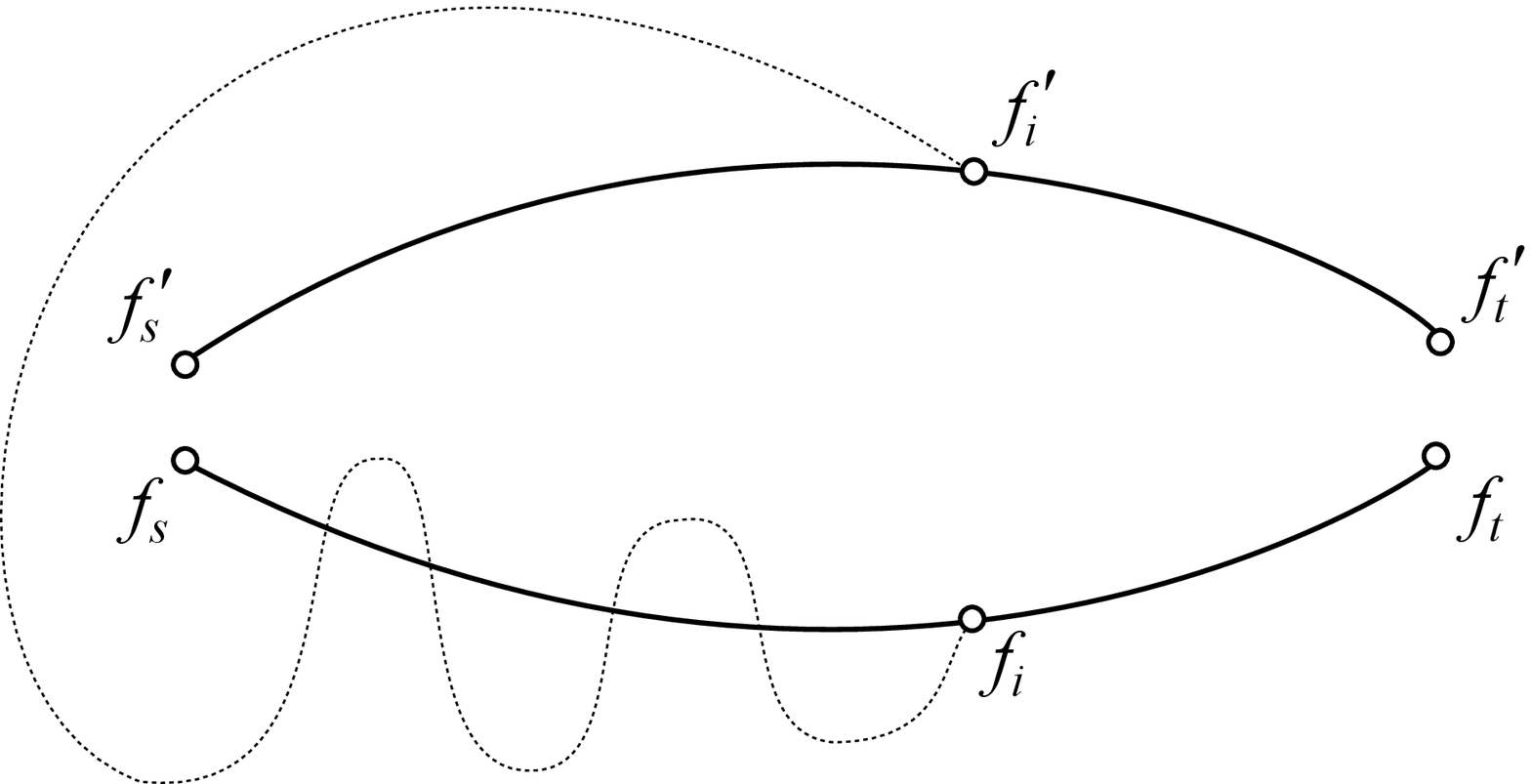,width=7cm}\\(b)\\
\end{tabular}
}
\vspace{5mm}
\centerline{
\begin{tabular}{c}
\psfig{file=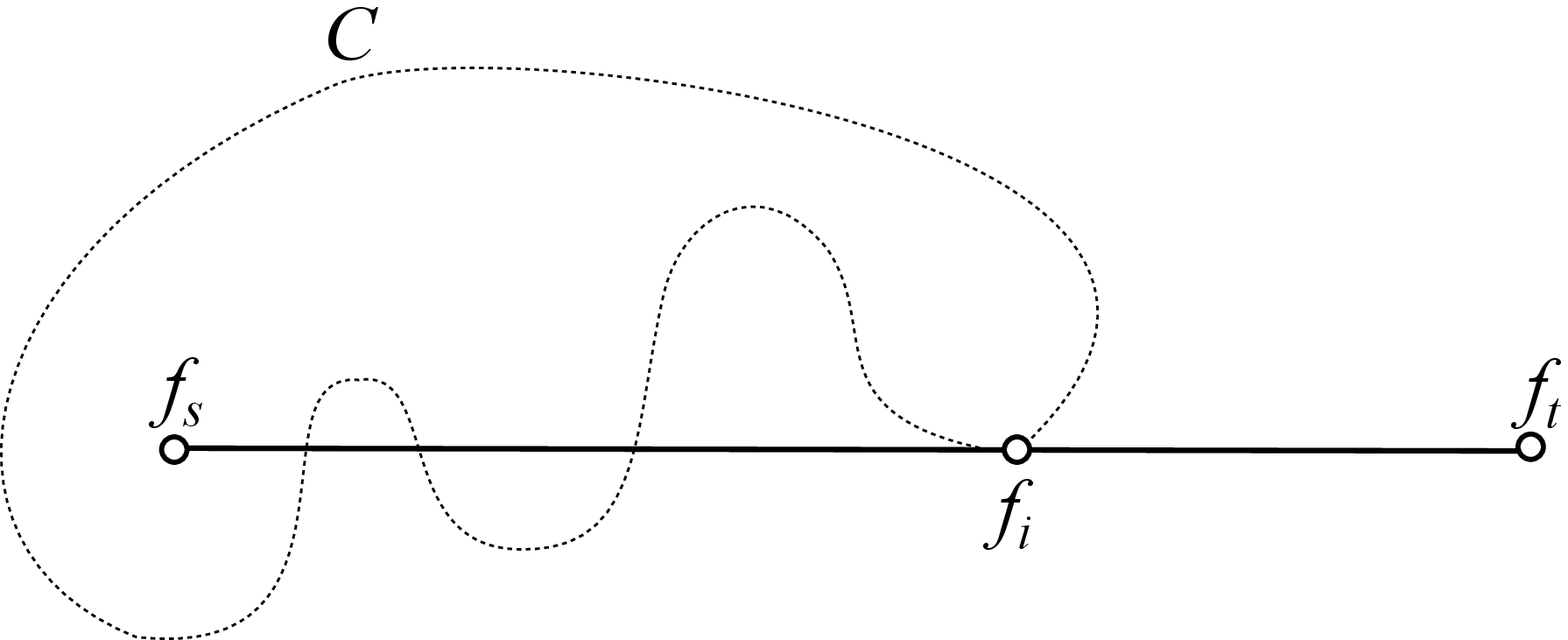,width=7cm}\\(c)\\
\end{tabular}
\hspace{7mm}
\begin{tabular}{c}
\psfig{file=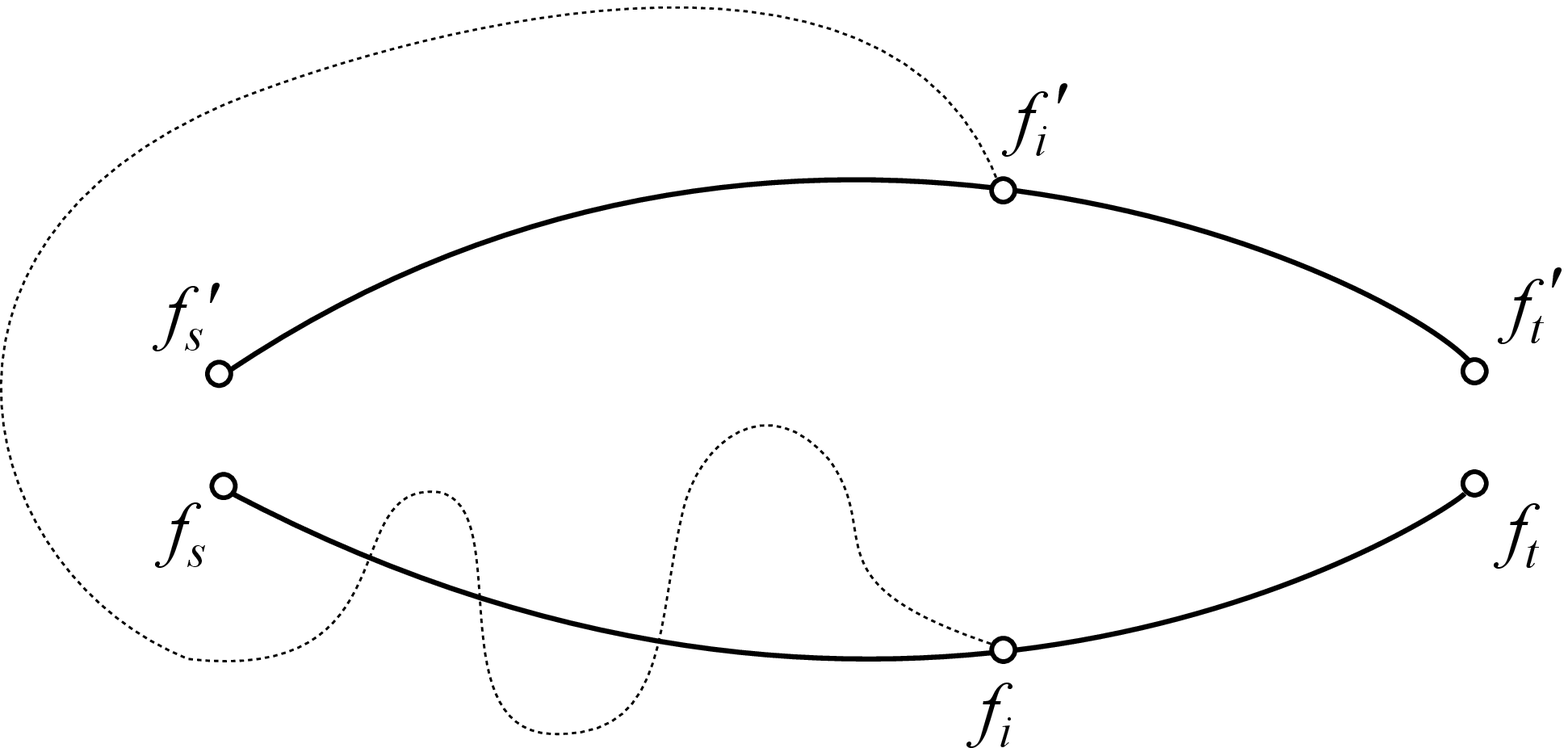,width=7cm}\\(d)\\
\end{tabular}
}
\vspace{5mm}
\centerline{
\begin{tabular}{c}
\psfig{file=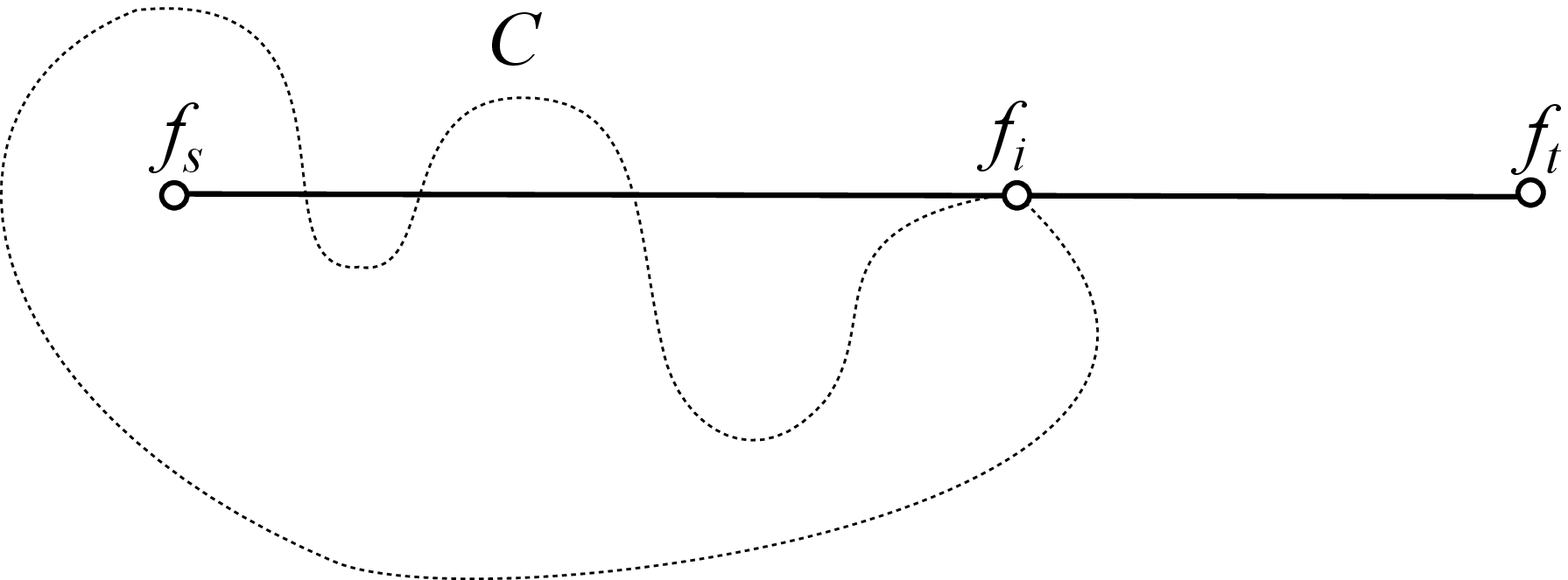,width=7cm}\\(e)\\
\end{tabular}
\hspace{7mm}
\begin{tabular}{c}
\psfig{file=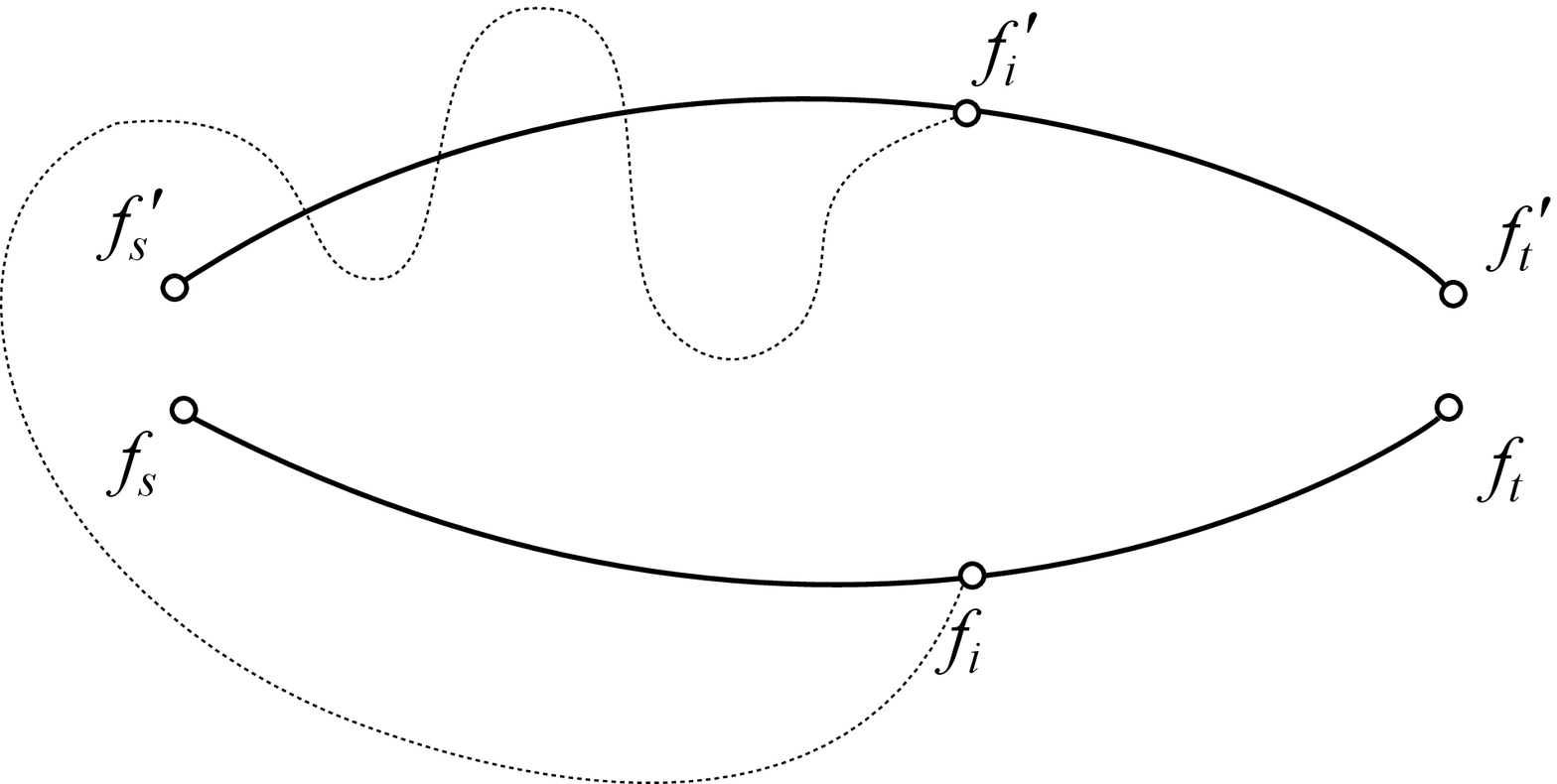,width=7cm}\\(f)\\
\end{tabular}
}
\caption{
On the proof of Lemma~\ref{lem:equivalence}. 
(a) The $f_i$-cut-cycle $C$ crosses the path $\pi$ in $f_i$ in the graph $G_D$.
(b) The corresponding $f_i$-separating path $\rho$ in $G_{\pi}$.
(c) The $f_i$-cut-cycle $C$ touches from above the path $\pi$ in $f_i$ in the graph $G_D$.
(d) The corresponding $f_i$-separating path $\rho$ in $G_{\pi}$.
(e) The $f_i$-cut-cycle $C$ touches from below the path $\pi$ in $f_i$ in the graph $G_D$.
(f) The corresponding $f_i$-separating path $\rho$ in $G_{\pi}$.
}
\label{fig:equivalence}
\end{figure}

\begin{lemma}
\label{lem:equivalence}
The cost of a minimum $f_i$-cut-cycle in $G_D$ is equal to the length of a shortest $f_i$-separating path in $G_{\pi}$.
\end{lemma}
\begin{proof}
Let $C$ be some $f_i$-cut-cycle in $G_D$: we show that there must be some $f_i$-separating path $\rho$ in $G_{\pi}$ having the same cost as $C$.
Note that the $f_i$-cut-cycle $C$ must either cross or touch the path $\pi$ in face $f_i$. First, assume that $C$ crosses $\pi$ in $f_i$.  Note that in this case $C$ has to cross the path $\pi$ an even number of times (excluding $f_i$), as otherwise $C$ would not separate $s$ from $t$ (see Figure~\ref{fig:equivalence}(a)). We can go along $C$ starting from $f_i$ in the graph $G_{\pi}$ and each time when $C$ crosses $\pi$ in $G_D$, we switch between $G_{\pi}'$ and $G_{\pi}''$  in $G_{\pi}$. Hence, due to parity of the number of crossings with $\pi$, this will produce a resulting path $\rho$ in $G_{\pi}$ which must end in $f_i'$ (see Figure~\ref{fig:equivalence}(b)). Second, assume that $C$ touches $\pi$ in $f_i$. Then $C$ has to cross path $\pi$ an odd number of times  (see Figure~\ref{fig:equivalence}(c)). Again if we trace $C$ starting from $f_i$ in $G_{\pi}$ we will produce a path $\rho$ in $G_{\pi}$ ending up in $f_i'$  (see Figure~\ref{fig:equivalence}(d)). Moreover, since $C$ is a $f_i$-cut-cycle in $G_D$, it cannot contain by definition any face $f_j$, for $j>i$: consequently, in either case the resulting path $\rho$ in $G_{\pi}$ will contain neither $f_j$ nor $f_j'$, for $j>i$.


Conversely, let $\rho$ be some $f_i$-separating path in $G_{\pi}$: we show that there must be some $f_i$-cut-cycle $C$ in $G_D$ having the same cost as $\rho$.
First, assume that $\rho$ enters $f_i$ and $f_i'$ using edges from the same graph, i.e., either $G_{\pi}'$ or $G_{\pi}''$. If this is the case, then the two edges must be such that one is from above and the other is from below.
By tracing $\rho$ in $G_D$ we obtain a cycle $C$ that crosses $\pi$ in $f_i$. This cycle does not need to be simple, but it has to cross $\pi$ an odd number of times (including the crossing at $f_i$), as each time when $C$ crosses $\pi$ the path $\rho$ must switch between $G_{\pi}'$ and $G_{\pi}''$. Hence, $C$ must be a cut-cycle.
Second, assume that
$\rho$ enters $f_i$ and $f_i'$ using edges from different graphs, i.e., one edge from $G_{\pi}'$ and the other edge from $G_{\pi}''$. If this is the case, then the two edges must be both from above or both from below. Consider first the case where $\rho$
leaves $f_i$ from above using an edge from $G_{\pi}''$, and enters $f_i'$ from above using an edge from $G_{\pi}'$ (see Figure~\ref{fig:equivalence}(d)). This time we get a cut-cycle $C$ touching $\pi$ in $f_i$ and intersecting $\pi$ an odd number of times. The case where $\rho$
leaves $f_i$ from below using an edge from $G_{\pi}'$, and enters $f_i'$ from below using an edge from $G_{\pi}''$ is completely analogous (see Figure~\ref{fig:equivalence}(f)).
Once again, since $\rho$ is an $f_i$-separating path in $G_{\pi}$, in any case the resulting cut cycle $C$ will not contain $f_j$, for $j>i$.
\end{proof}

Note that, having constructed the graph $G_{\pi}$, we can find the shortest $f_i$-separating path by simply running Dijkstra's algorithm on $G_{\pi}$ where we remove all faces $f_j$ and $f_j'$ for $j>i$. By Lemmas~\ref{lem:Johnson}, \ref{lem:minimum-cut-cycle} and~\ref{lem:equivalence},
the linear-time implementation of Dijkstra's algorithm known for planar graphs~\cite{henzinger-et-al-97} implies an $O(n^2)$ time algorithm for computing minimum cuts in planar graphs.
There is a more efficient way of computing minimum cuts that goes along the lines of Reif's recursive algorithm~\cite{Reif83}. Before describing this approach, we need to prove some non-crossing properties of $f_i$-cut-cycles.

\begin{figure}[htp]
\centerline{\psfig{file=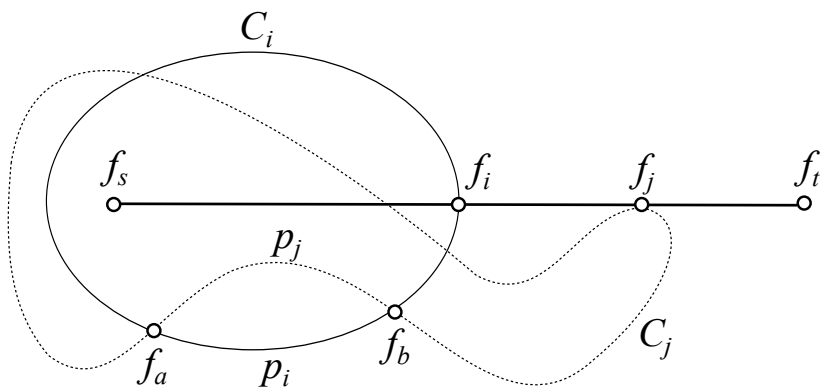,height=5cm}}
\caption{On the proof of Lemma~\ref{lemma-cut-cycle-min}.}
\label{fig:non-crossing}
\end{figure}

\begin{lemma}
\label{lemma-cut-cycle-min}
For $i<j$, let $C_i$ be a minimum $f_i$-cut-cycle in $G_D$, and let $C_j$ be a minimum $f_j$-cut-cycle in $G_D$. Then there exists
a cut cycle $C \subseteq \intc(C_i)\cap \intc(C_j)$ in $G_D$ such that $c(C) \le c(C_i)$.
\end{lemma}
\begin{proof}
If $C_i \subseteq \intc(C_j)$ then $C=C_i$ and the lemma follows trivially. On the other hand, it is impossible that
$C_j \subseteq \intc(C_i)$ because in this case $C_i$ would include some face $f_k$ on $\pi$ for $k\ge j>i$.

The only possibility left is that $C_i \nsubseteq \intc(C_j)$ and $C_j \nsubseteq \intc(C_i)$. In this case, there must exist a subpath $p_i$ of $C_i$
from $f_a$ to $f_b$ such that $p_i$ intersects $\intc(C_j)$ only at $f_a$ and $f_b$.
Let $p_j$ be the subpath of $C_j$ going from $f_a$ to $f_b$
(see Figure~\ref{fig:non-crossing}). We claim  that $c(p_j) \le c(p_i)$.
Indeed, suppose by contradiction that $c(p_j) > c(p_i)$, and let $C_j'$ be the cycle obtained from $C_j$ after replacing path $p_j$ with $p_i$.
Then the cycle $C_j'$ is shorter than $C_j$. Moreover, since $C_i$ does not include any face $f_k$ for $k>i$, also $C_j'$ cannot include any face $f_k$ for $k>j>i$. As a consequence, the cycle $C_j'$ is an $f_j$-cut-cycle in $G_D$, with $c(C_j')<c(C_j)$, contradicting our assumption that $C_j$ is a minimum $f_j$-cut-cycle in $G_D$.

Since $c(p_j) \le c(p_i)$,  replacing path $p_i$ on $C_i$ with the path $p_j$ yields  a cycle $C_i'$, with $c(C_i')\leq  c(C_i)$. As long as $C_i' \nsubseteq \intc(C_j)$ and $C_j \nsubseteq \intc(C_i')$ we can repeat this procedure. At the end, we will obtain a cycle $C \subseteq \intc(C_i)\cap \intc(C_j)$, such that $c(C) \le c(C_i)$.
Moreover, the obtained cycle $C$ will be a cut-cycle, as $\intc(C_i) \cap \intc(C_j)$ contains $s$.
\end{proof}

\begin{lemma}
\label{lemma-cut-cycle-int-min}
For $i<j$, let $C_i$ be a minimum $f_i$-cut-cycle in $G_D$, and let $C_j$ be a minimum $f_j$-cut-cycle in $G_D$. Then, for some $i'\le i$,
there exists a minimum $f_{i'}$-cut-cycle $C_{i'}\subseteq \intc(C_j)$ such that $c(C_{i'})\le c(C_i)$.
\end{lemma}
\begin{proof}
Consider the cut-cycle $C$ contained inside $\intc(C_i)\cap \intc(C_j)$ as given by Lemma~\ref{lemma-cut-cycle-min}, for which we know that $c(C) \le c(C_i)$. As $C \subseteq \intc(C_i)$,  $C$ cannot
contain any $f_k$ such that $k>i$. Hence, it is an $f_{i'}$-cut-cycle for some $i'\le i$. The minimum $f_{i'}$-cut-cycle $C_{i'}$ is
shorter than $C$ and consequently it is shorter than $C_i$. If $C_{i'}\subseteq\intc(C_j)$, the lemma follows. Otherwise, we can apply Lemma~\ref{lemma-cut-cycle-min} to produce another minimum $f_{i''}$-cut-cycle contained in $\intc(C_j)$ such that $c(C_{i''})\le c(C_i)$, for some $i''\le i'\le i$.
\end{proof}

The above lemma shows that each computed cut splits the graph into two parts, the interior and the exterior part of the cut, which can be handled separately. Hence, we can use a divide and conquer approach on the path $\pi$. We first find a minimum cut-cycle that contains the middle vertex on the path $\pi$. Then we recurse on both parts of the path, so there will be $O(\log n)$ levels of recursion in total. In this recursion we will compute minimum cost $s$-$t$ cuts for all vertices on $\pi$. By Lemma~\ref{lem:equivalence}, we compute a minimum $f_i$-cut-cycle for some $f_i$ by finding
a shortest path in the planar graph $G_{\pi}$. Then we need to divide the graph into the inside and the outside of the cut. The vertices on the minimum cut need to be included into both parts, so we need to take care that the total size of the parts does not increase too much. This can be done in a standard way as described by Reif~\cite{Reif83} or by Hassin and Johnson~\cite{hassin-johnson-85}. Their technique guarantees that on each level of the recursion the total size of the parts is bounded by $O(n)$. Hence, using the $O(n)$-time algorithm~\cite{henzinger-et-al-97} for shortest paths we get an $O(n\log n)$ time algorithm for finding minimum cuts in planar graphs. Although this approach yields the same time bounds as previously known algorithms~\cite{hassin-johnson-85,Reif83}, it has the main advantage to allow the use of any path $\pi$ in $G_D$, while previous algorithms  were constrained to choose $\pi$ as a shortest path. As shown in the next section, we can even allow the path $\pi$ to be non-simple. In Section~\ref{section-cluster} we will show how to exploit the freedom implicit in the choice of the path $\pi$ to produce a faster $O(n\log\log n)$ time algorithm for finding minimum cuts in planar graphs.

\subsection{Using Non-simple Paths}
\label{section-non-simple-paths}
Let $\pi=(v_1, \ldots, v_{\ell})$ be a non-simple path and let $v$ be a vertex appearing at least twice on the path $\pi$, i.e., $v=v_i=v_j$ for some $i<j$. We say that the path $\pi$ is {\it self-crossing} in $v$ if the edges incident to $v$ on $\pi$ appear in the following circular order $(v_{i-1},v_i)$, $(v_{j-1},v_j)$, $(v_i,v_{i+1})$, $(v_j,v_{j+1})$ in the embedding. We say that a path $\pi$ is {self-crossing} if $\pi$ is self-crossing in some vertex $v$. Otherwise we say that $\pi$ is {\it non-crossing}. If a vertex $v$ appears at least twice on a non-crossing path $\pi$ than we say that $\pi$ {\it touches itself} in $v$. In the previous section we assumed that the path $\pi$ from $f_s$ to $f_t$ in $G_D$ was simple, now we will show that we only need the weaker assumption that $\pi$ is non-crossing.

\begin{figure}[thp]
\centerline{
\hspace{3cm}
\begin{tabular}{c}
\psfig{file=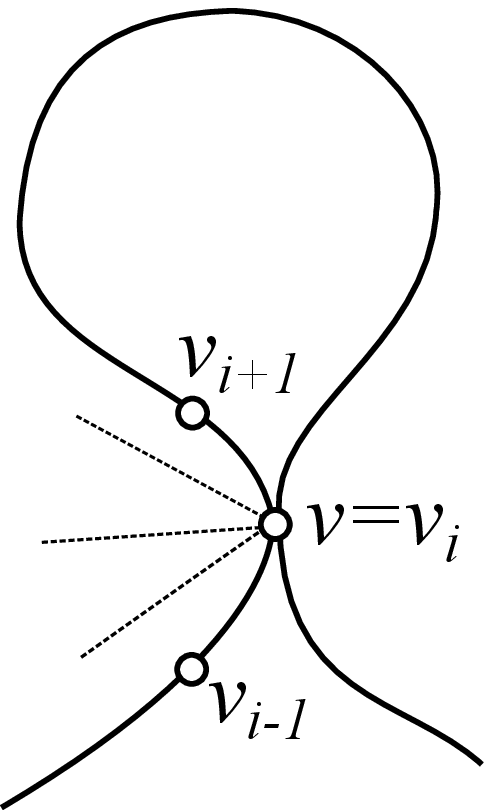,height=4cm}\\(a)\\
\end{tabular}
\hspace{3cm}
\begin{tabular}{c}
\psfig{file=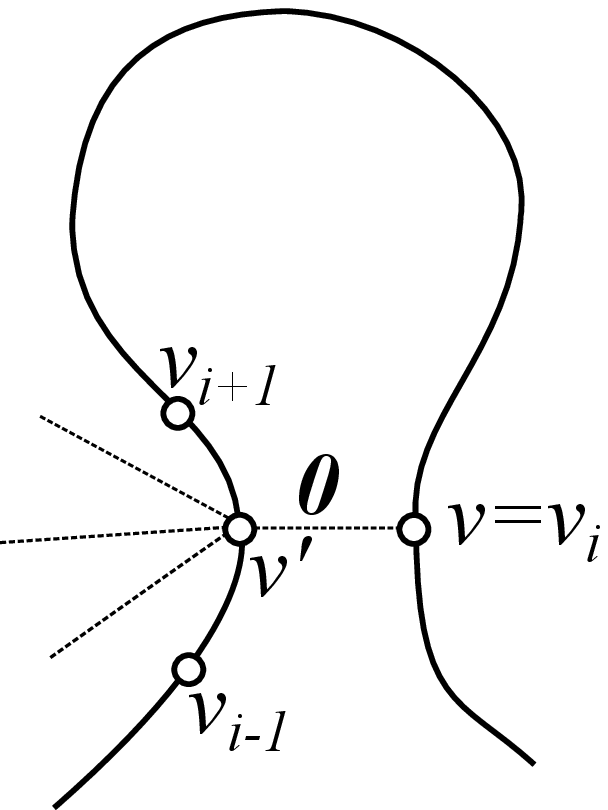,height=4cm}\\(b)\\
\end{tabular}
\hspace{3cm}
}
\caption{
Dealing with non-simple paths. 
(a) A non-crossing path $\pi$ of $G_D$ that touches itself at face $v=v_i$.
(b) The vertex splitting transformation on face $v$.
}
\label{fig:corollary1}
\end{figure}

In order to work with non-crossing paths we will modify the graph $G_D$ to make the path $\pi$ simple. Let $v=v_i$ be a face where $\pi$ touches itself in $G_D$. Note that there is no other edge from $\pi$ between the edges $(v_{i-1},v_i)$, $(v_i,v_{i+1})$ in the circular order around $v$ given by the embedding. Take all edges $E_v$ incident to $v$ that are embedded between and including the edges $(v_{i-1},v_i)$, $(v_i,v_{i+1})$. Now, add a new face $v'$ to $G_D$ and make the edges $E_v$ to be incident with $v'$ instead of $v$. Finally connect $v$ with $v'$ using an undirected edge of length zero
(see Figure~\ref{fig:corollary1}). Let $\pi$ be a non-crossing path:
perform this vertex-splitting operation until $\pi$ becomes a simple path. This produced a new graph $G_{D,\pi}$. Note that this
transformation
does not change the lengths of cut-cycles, so we get the following observation.

\begin{corollary}
The lengths of minimum $f_i$-cut-cycles in $G_D$ and $G_{D,\pi}$ are the same.
\end{corollary}

As a result, if $\pi$ is a non-simple non-crossing path,  instead of running our algorithm on $G_D$, we can compute the graph $G_{D,\pi}$ and run our algorithm on $G_{D,\pi}$. 
\section{Cluster Partitions}
\label{section-cluster}

Our algorithms are based on a particular cluster decomposition.
We start by presenting the ideas behind this cluster decomposition and then show how this decomposition  can be effectively exploited for computing min-cuts.
From now on, we assume that we are given a graph for which the dual graph has vertex degree at most three. This is without loss of generality, since it can be obtained by triangulating the primal graph with zero capacity edges.

Let $n$ be the number of vertices of $G$. We first define a {\it cluster partition} of $G$ into edge clusters which will be used by our algorithm.
In the cluster partition the graph is decomposed into a set of cluster $\mathcal{P}$ such that each edge of $G$ is included into exactly one cluster. 
Each cluster contains two types of vertices: {\em internal} vertices, and {\em border} vertices. An internal vertex is adjacent only to vertices in the same cluster, while a border vertex is adjacent to vertices in different clusters.
A {\it hole} in a cluster is any face (including the external face) containing only boundary vertices. We denote by $\partial P$ the set of border vertices of a cluster $P$.
We define an {\it $r$-partition} of an $n$-node graph to be a cluster partition that contains at most $O(\frac{n}{r})$ clusters, each
containing at most $r$ vertices, $O(\sqrt{r}\,)$ border vertices and a constant number of holes. The proof of the following lemma is in Appendix~\ref{app-clusters}. It is based on ideas of Frederickson~\cite{federickson-87}, who constructed a similar partition without the bound on the number of holes.
\begin{lemma}
\label{lemma-computing-partition}
An $r$-partition of an $n$-node planar graph can be computed in $O(n \log r + \frac{n}{\sqrt{r}} \log n)$ time.
\end{lemma}

We use the $r$-partition to define a representation for shortest paths in a graph that has similar number of edges, but fewer vertices. In order to achieve this, we use the notion of dense distance graphs. A {\it dense distance graph} for a cluster $C$ is defined to be a complete graph over the border vertices of $C$ where edge lengths correspond to shortest path distances in $C$. In order to compute dense distance graphs for all clusters we use Klein's algorithm~\cite{klein-05}, who have shown that after $O(n\log n)$ preprocessing time any distance from the external face can be computed in $O(\log n)$ time. The proof of the following lemma is given in Appendix~\ref{appendix-computing-dd}.

\begin{lemma}
\label{lemma-computing-dd}
Given an $r$-partition $\mathcal{P}$ we can compute a dense distance graph for all clusters in $\mathcal{P}$ in $O(n \log r)$ time.
\end{lemma}

The dense distance graphs can be used to speed up shortest path computations using Dijkstra's algorithm. It was shown by Fakcharoenphol and Rao (\cite{fr06}, Section 3.2.2) that a Dijkstra-like algorithm can be executed on a dense distance graph for a cluster $P$ in $O(|\partial P| \log^2 |P|)$ time.  Having constructed the dense distance graphs, we can run Dijkstra in time almost proportional to the number of vertices (rather than to the number of edges, as in standard Dijkstra). We use this algorithm in graphs composed of dense distance graphs and a subset $E'$ of edges of the original graph $G=(V,E)$:

\begin{corollary}\label{cor:hybrid-FR}
  Dijkstra can implemented in $O(|E'| \log |V|  +  \sum_{i}|\partial G_i|\log^2 |\partial G_i|)$ time on a graph composed of a set of dense distance graphs $G_i$ and a set of edges $E'$
  over the vertex set $V$.
\end{corollary}
\begin{proof}
In order to achieve this running time we use Fakcharoenphol and Rao~\cite{fr06} data-structure for each $G_i$. Moreover, minimum distance vertices from each $G_i$ and all endpoints of edges in $E'$ are kept in a global heap.
\end{proof}

In general, clusters may contain holes although in typical cases, e.g., in grid graphs, the obtained clusters are holeless. In this section, in order to introduce the main ideas behind our algorithm, we restrict ourselves to {\it holeless $r$-partitions}, i.e., $r$-partitions where each cluster contains one hole (external face). We will show in Appendix~\ref{appendix-general-clusters} how to handle the general case.

Assume that we have computed the dense distance graphs for all clusters in a given $r$-partition $\mathcal{P}$ of the dual graph $G_D$. Recall that in our min-cut algorithm we are free to choose any path $\pi$ from $f_s$ to $f_t$ in $G_D$, as long as $\pi$ is non-crossing. We choose $\pi$ to minimize the number of clusters it crosses and to maximally use the dense representation of clusters.

We define a {\it skeleton graph} $G_{\mathcal{P}}=(\partial \mathcal{P}, E_{\mathcal{P}})$ to be a graph over the set of border vertices in $\mathcal{P}$. The edge set $E_{\mathcal{P}}$ is composed of infinite length edges connecting consecutive (in the order on the hole) border vertices on each hole. By our holeless assumption, all border vertices in each cluster lie on the external face of the cluster, so the graph $G_{\mathcal{P}}$ is connected. We define a {\it patched graph} to be $\overline{G} = G_D \cup G_{\mathcal{P}}$. Note that this graph is still planar and the shortest distances do not change after adding infinite length edges.

\begin{figure}[htp]
\centerline{\psfig{file=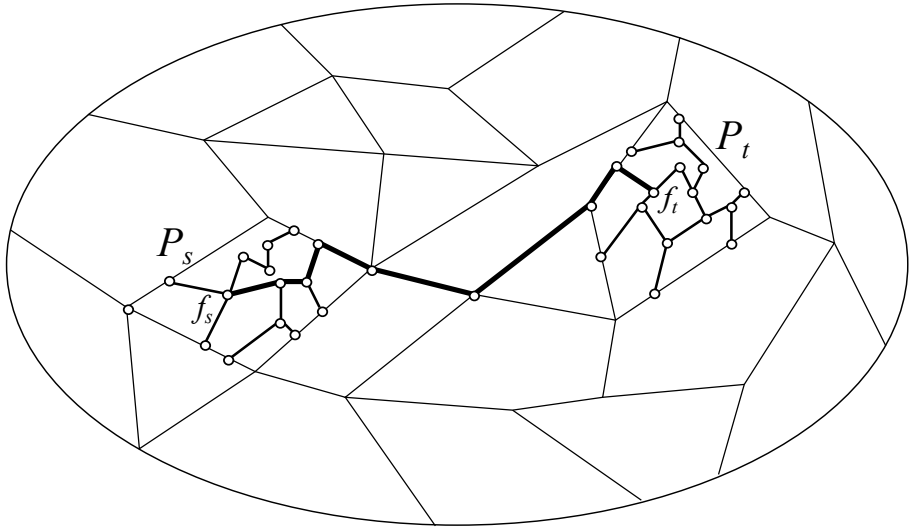,width=12cm}}
\caption{The graph $\overline{G}^{s,t}$ containing clusters $P_s$ and $P_t$, the dense distance graphs for all other clusters, and the skeleton graph $G_{\mathcal{P}}$. The path $\pi$ from $f_s$ to $f_t$ is shown in bold.}
\label{fig:Gsignedst}
\end{figure}

Define $\overline{G}^{s,t}$ to be the graph composed of:
(1) two clusters $P_s$ and $P_t$ that include $f_s$ and $f_t$ respectively;
(2) the dense distance graphs (represented by square matrices) for all other clusters;
(3) the skeleton graph $G_{\mathcal{P}}$ (see Figure~\ref{fig:Gsignedst}).
Note that $\overline{G}^{s,t}$ contains:
$O(2r + \frac{n}{r}\sqrt{r}\,) = O(r + \frac{n}{\sqrt{r}})$ vertices;
$O(\frac{n}{r})$ dense distance graphs each over $O(\sqrt{r}\,)$ vertices;
at most $3r$ edges of $G_D$ from $P_s$ and $P_t$; additional $O(\frac{n}{\sqrt{r}})$ edges from the skeleton graph $G_{\mathcal{P}}$.
Using Corollary~\ref{cor:hybrid-FR} to run Dijkstra's algorithm we get the following. 


\begin{corollary}
\label{cor-fast-shortest-paths}
The shortest paths in $\overline{G}^{s,t}$ can be computed in $O((r+\frac{n}{\sqrt{r}})\log^2 n)$ time.
\end{corollary}

\subsection{Recursive Division}
Let $b_s$ and $b_t$ be any border vertices in clusters $P_s$ and $P_t$ respectively. We define $\pi$ to be composed of:
a simple path from $f_s$ to $b_s$ in $P_s$;
a simple path from $b_s$ to $b_t$ in $G_{\mathcal{P}}$;
and a simple path from $b_t$ to $f_t$ in $P_t$.
Note that the construction of the graph $\overline{G}^{s,t}$ and consequently of $\overline{G}^{s,t}_{\pi}$ takes $O(n \log r + \frac{n}{\sqrt{r}} \log n)$ time by Lemmas~\ref{lemma-computing-partition} and~\ref{lemma-computing-dd}.

\begin{figure}[htp]
\centerline{\psfig{file=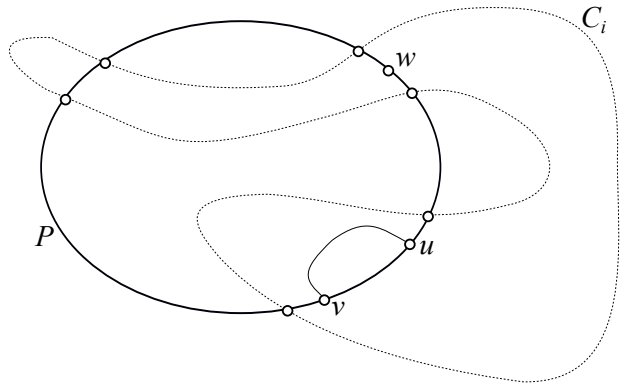,height=5cm}}
\caption{Computing $G_P \cap \intc(C_i)$. It two border vertices $u$ and $v$ are not separated by the $f_i$-cut-cycle $C_i$, then their distance in $G_P\cap \intc(C_i)$ is equal to their distance in $G_P$. On the other hand, if border vertices $v$ and $w$ are separated by the $f_i$-cut-cycle $C_i$, then their distance in $G_P\cap \intc(C_i)$ is infinite.}
\label{fig:section3}
\end{figure}

Let $C_i$ be some $f_i$-cut-cycle in $G_D$. After finding $C_i$, we need to recurse on the graphs $G_D \cap \intc(C_i)$ and $G_D  \cap \extc(C_i)$. These graphs cannot be computed explicitly. However, we will show how to determine $\overline{G}^{<i} := \overline{G_D  \cap \intc{(C_i)}}^{s,t}$ and $\overline{G}^{i>} :=\overline{G_D  \cap \extc{(C_i)}}^{s,t}$ using $\overline{G}^{s,t}$. Let $P$ be a cluster in the partition other then $P_s$ or $P_t$, and let $G_{P}$ be its dense distance graph. For a set of vertices $X$ we define $G_P \cap X$ to be the dense distance graph of $P \cap X$. The $f_i$-cut-cycle was found using $\overline{G}^{s,t}$ so parts of $C_i$ that pass through $P$ correspond to shortest paths. Hence, the shortest paths in $P \cap \intc(C_i)$ and $P \cap \extc(C_i)$ cannot cross the cycle $C_i$. As a result, distances in $G_P \cap \intc(C_i)$ and $G_P \cap \extc(C_i)$ between border vertices of $P$
that are not separated by $C_i$ are the same as in $G_P$. On the other hand, for border vertices that are separated by $C_i$ the distances are infinite (see Figure~\ref{fig:section3}).
We define $\overline{G}^{s,t} \cap X$ to be the graph obtained by taking $G_P \cap X$ for every cluster in $G$. Note that we have the following.
\begin{corollary}
\label{cor-recursive-skeletons}
$\overline{G}^{<i} = \overline{G}^{s,t} \cap \intc(C_i)$ and $\overline{G}^{i>} = \overline{G}^{s,t} \cap \extc(C_i)$.
\end{corollary}

Using these equalities the construction of $\overline{G}^{<i}$ and $\overline{G}^{i>}$ takes time proportional to the size of $\overline{G}^{s,t}$ only.
Given a path $\pi = f_1,\ldots,f_k$, the recursive algorithm for computing min-cuts works as follows.
First, remove from $\overline{G}^{s,t}$ vertices with degree $2$ by merging the two incident edges,
and find an $f_i$-cut-cycle $C_i$ for $i = \lfloor \frac{k}{2} \rfloor$. Next,
construct graphs $\overline{G}^{<i}$ and $\overline{G}^{i>}$, and
recursively find a minimum cut $C_{<i}$ in $\overline{G}^{<i}$ and a minimum cut $C_{i>}$ in $\overline{G}^{i>}$. Finally, return the smallest of the three cuts $C_{<i}$, $C_{i}$ and $C_{i>}$.

To achieve our promised bounds, we need to show that the above algorithm works in sublinear time. In order to show that, we only need to prove that the total size of the graphs on each level in the recursion tree is small.
Take a graph $\overline{G}^{s,t}$ and for each cluster $P$ that contains more then one border vertex add a new vertex $v_P$ and replace each dense distance graph by a star graph with $v_P$ in a center. We call the resulting graph the {\it contracted graph} and denote it by $\overline{G}^{s,t}_c$.
Obviously, the contracted graph has more vertices than the original graph: thus, we can bound the number of vertices in the original graph by considering only contracted graphs.
The proof of this lemma is included in Appendix~\ref{appendix-size-of-graphs}.

\begin{lemma}
\label{lemma-size-of-graphs}
The total number of vertices in contracted graphs on each level in the recursion tree is $O(r + \frac{n}{\sqrt{r}})$.
\end{lemma}

By the above lemma and  Corollary~\ref{cor:hybrid-FR}, running Dijkstra for each level takes $O((r+\frac{n}{\sqrt{r}})\log^2 n)$ time in total. On each level the length of the path $\pi$ is halved, so there are at most $\log n$ recursion levels. Hence we obtain the following theorem.

\begin{theorem}
\label{theorem-holeless-min-cuts}
Let $G$ be a flow network with the source $s$ and the sink $t$. If $G_D$ allows holeless $r$-partition, then the minimum cut between $s$ and $t$ can be computed in $O(n \log r + (r+\frac{n}{\sqrt{r}})\log^3 n)$ time. By setting $r=\log^8 n$ we obtain an $O(n \log \log n)$ time algorithm. 
\end{theorem}

Theorem~\ref{theorem-holeless-min-cuts} holds also for general $r$-partitions within the same  $O(n \log \log n)$ time bound.
The modifications needed to make the above algorithm work in the general case are presented in Appendix~\ref{appendix-general-clusters}. 
\section{Computing Maximum Flows}
\label{section-max-flow}
The standard ways of computing maximum flow in near linear time assume that we already have computed its flow value $f$. It is given by the
min-cut capacity and as we already know it can be computed in $O(n \log \log n)$ time. We will use the approach proposed by Hassin and Johnson ~\cite{hassin-johnson-85}, but adopted to our case as we use a different family of cuts $C_i$. As argued in the following, their approach uses only very basic properties of the cut-cycles and can be directly applied to our case. Moreover, for the sake of brevity, we will assume that that we are given a holeless $r$-partition. The general case can be handled using ideas presented in Appendix~\ref{appendix-general-clusters}, and it will be included in the full version of this paper.

Let us define the graph $\overrightarrow{G}_{\pi}$ to be the graph obtained
from $G_{\pi}'$ by adding directed edges of length $-f$ from $f_i$ to $f_i'$ for all $1\le i \le k$. After Lemma~4.1 in Miller and Naor~\cite{miller-naor-89} we know the following.

\begin{corollary}
The graph $\overrightarrow{G}_{\pi}$ does not contain negative length cycles.
\end{corollary}

Let $r$ be such that $C_r$ is the shortest of all $C_i$ for $1\le i \le k$. The above corollary assures that distance $\dist(v)$ from $f_r'$ to a vertex $v$ is well defined in $\overrightarrow{G}_{\pi}$. The next lemma follows by Theorem~1 in~\cite{hassin-johnson-85} or Section~5.1 in~\cite{miller-naor-89}.

\begin{lemma}
\label{lemma-flow-from-dual}
Let $e = (u,v)$ be the edge in $G$ and let $e_D = (f_u, f_v)$ be the corresponding edge in $G_D$. The face $f_u$ is
defined to lie to the left when going from $u$ to $v$. The function $f(u,v) := \dist(f_u) - \dist(f_v)$ defines the maximum
flow in $G$.
\end{lemma}

By this lemma, in order to construct the flow function we only need to compute distances from $f_r'$ in $\overrightarrow{G}_{\pi}$. The cycles $C_i$ for $i=1,\ldots, k$ divide $\overrightarrow{G}_{\pi}$ into $k+1$ subnetworks $\overrightarrow{G}_0,\ldots, \overrightarrow{G}_k$ where for $1 \le i \le k-1$, $\overrightarrow{G}_i$ is the subnetwork bounded by and including $C_i$ and $C_{i+1}$.  The next lemma is a restatement of Lemma~2 from~\cite{hassin-johnson-85}.

\begin{lemma}
\label{lemma:restrict}
Let $v$ be a vertex in $N_i$. Then if $i<r$ then there exists a shortest $(f_r',v)$-path which is contained in $\bigcup_{j=0}^{r-1} \overrightarrow{G}_j$. Similarly, if $i\ge r$ then there exists a shortest $(f_r',v)$-path which is contained in $\bigcup_{j=r}^{k} \overrightarrow{G}_j$.
\end{lemma}

Lemma~\ref{lemma:restrict} implies that the computation of $\dist(v)$ for $v \in \bigcup_{j=0}^{r-1} \overrightarrow{G}_j$ can be restricted to $\bigcup_{j=0}^{r-1} \overrightarrow{G}_j$ only. We can restrict the computation for $v \in \bigcup_{j=r}^{k} \overrightarrow{G}_j$ in a similar fashion.

Similarly to~\cite{hassin-johnson-85} let us define a {\it normal} path to be a simple $(f_r', v)$-path $\rho(v) = \rho_r\cdot \ldots \cdot \rho_q \cdot \ldots \cdot \rho_{2q-i}$ such that, for $j=r, \ldots, q$, subpath $\rho_j$ is in $\overrightarrow{G}_j$, and, for $j=q+1, \ldots, 2q-i$, subpath $\rho_j$
is in $\overrightarrow{G}_{2g-j}$ and uses no edges of negative length. We require as well that $q$ is minimal. Hassin and Johnson~\cite{hassin-johnson-85}
have shown that distances from $f_r'$ in an $n$-vertex graph can be computed in $O(n \log n)$ time when for each vertex $v$ there exists a shortest $(f_r',v)$-path that is normal (Theorem~2 in~\cite{hassin-johnson-85}). Their computation is based on Dijkstra's algorithm and can be sped up using dense distance graphs and Corollary~\ref{cor:hybrid-FR}.

\begin{lemma}
\label{lemma-flow-normal}
Distances from $f_r'$ in the graph $\overrightarrow{G}_{\pi}$ can be computed in $O(n \log r + (\frac{n}{\sqrt{r}} + r)\log^2 n)$ time
when for each vertex $v$ there exists a shortest $(f_r',v)$-path that is normal.
\end{lemma}
\begin{proof}
The algorithm works in two phases. First, we take $\overrightarrow{G}_{\pi}$ and substitute each cluster $P$ with its dense distance graph. The resulting graph has $O(\frac{n}{\sqrt{r}})$ vertices. The Dijkstra-like computation from~\cite{hassin-johnson-85} can be executed in $O((\frac{n}{\sqrt{r}} + r)\log^2 n)$ time on this graph. In this way we obtain distances $\dist(v)$ for border nodes $v$ in all clusters in $G_D$. Second, for each cluster separately we run Hassin and Johnson's computation starting from border vertices only. The second phase works in $O(r \log r)$ time for each cluster, which yields $O(n \log r)$ time in total.
\end{proof}

In order to use the above lemma we only need the following result, which can be proven in the same way as Lemma~3 from~\cite{hassin-johnson-85}. The proof in~\cite{hassin-johnson-85} uses only the fact that subpaths of cut cycles $C_i$ are shortest paths and
this holds in our case as well.

\begin{figure}[thp]
\centerline{
\begin{tabular}{c}
\psfig{file=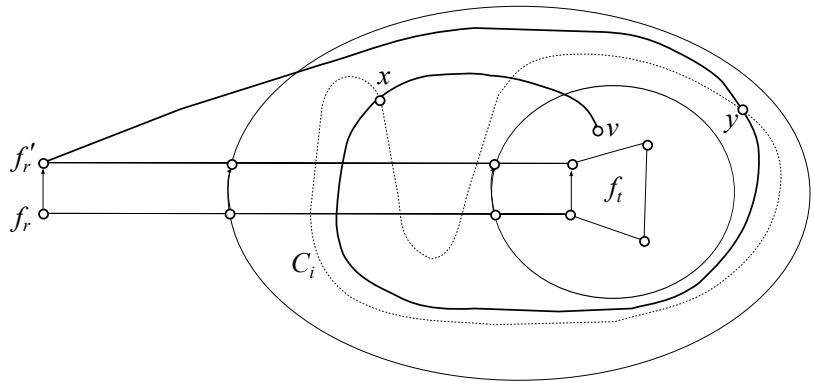,height=5cm}\\(a)\\
\end{tabular}
}
\vspace{7mm}
\centerline{
\begin{tabular}{c}
\psfig{file=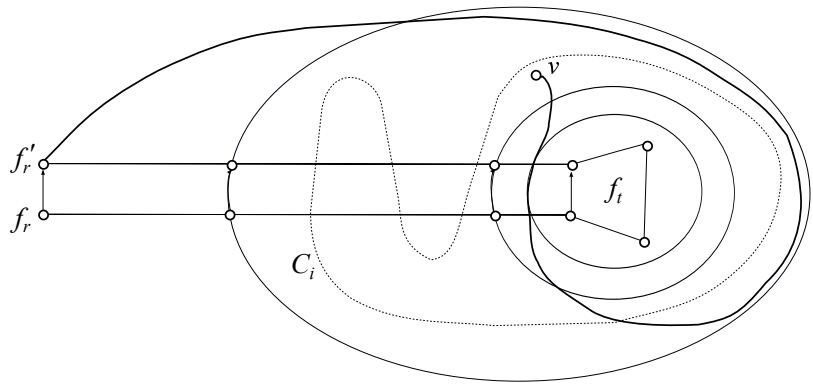,height=5cm}\\(b)\\
\end{tabular}
}
\caption{On the proof of Lemma~\ref{lemma-paths-normal}.
(a) The cycle $C_i$ can be short cut using the subpath between $x$ and $y$ in $\rho(v)$.
(b) A valid normal path.
}
\label{fig:lemma12}
\end{figure}

\begin{lemma}
\label{lemma-paths-normal}
For any $i=r, \ldots, k$, for every vertex $v$, in $\overrightarrow{G}_i$ there exists a normal shortest path $\rho(v)$.
\end{lemma}
\begin{proof}
Assume that a shortest path $\rho(v)$ intersect some cycle $C_i$. If the previous intersected cycle was $C_i$ as well, then
we can either short cut $\rho(v)$ using the part of $C_i$ or short cut $C_i$ using part of $\rho(v)$ (see Figure~\ref{fig:lemma12}). 
This contradicts either the minimality of $\rho(v)$ or the minimality of $C_i$. Moreover, after $\rho(v)$ leaves $C_{i}$
to go into $\overrightarrow{G}_{i-1}$ it cannot use a negative edge any more, as otherwise it would cross itself. If it crosses itself
then the resulting cycle could be removed as it cannot have a negative weight.
\end{proof}

Combining Lemma~\ref{lemma-flow-from-dual}, Lemma~\ref{lemma-flow-normal} and Lemma~\ref{lemma-paths-normal} we obtain the main result of
this section.

\begin{theorem}
The maximum flow in an undirected planar graph can be computed in $O(n \log r + (\frac{n}{\sqrt{r}} + r)\log^2 n)$ time. By setting $r=\log^8 n$ we obtain an $O(n \log \log n)$ time algorithm.
\end{theorem}

\section{Dynamic Shortest Paths and Max Flows in Planar Graphs}

Most of the ideas presented in this section are not entirely new, but nevertheless combined together they are able to simplify and improve previously known approaches. Our dynamic algorithm builds upon the $r$-partition introduced in Section~\ref{section-cluster}. We first show how to maintain a planar graph $G_D$ with positive edge weights under an intermixed sequence of the following operations~\footnote{We have chosen to work with the dual graph $G_D$ for consistency with the remaining parts of the paper.}:

\begin{mylist}{\em shortest\_path$(x,y)$  :}
\litem{\em insert$_{D}(x,y,c)$}  add to $G_D$ an edge of length $c$ between vertex $x$ and vertex $y$, provided that the new edge preserves the planarity of $G_D$;
\litem{\em delete$_D(x,y)$}  delete from $G_D$ the edge between vertex $x$ and vertex $y$;
\litem{\em shortest\_path$_D(x,y)$}  return a shortest path in $G_D$ from vertex $x$ to vertex $y$.
\end{mylist}

In our dynamic algorithm we maintain the $r$-partition of $G_D$ together with dense distance graphs for all clusters in the partition. This information will be recomputed every $\sqrt{r}$ operations. We now show how to handle the different operations. We start with operation {\em insert\/}: let $(x,y)$ be the edge to be inserted, and let ${P}_x$ (respectively ${P}_y$) be the cluster containing vertex $x$ (respectively vertex $y$). If $x$ and $y$ are not already border vertices in clusters ${P}_x$ and ${P}_y$, we make them border vertices in both clusters and add edge $(x,y)$ arbitrarily either to cluster ${P}_x$ or to cluster ${P}_y$. Next, we recompute the dense distance graphs of clusters
${P}_x$ and ${P}_y$, as explained in Lemma~\ref{lemma-computing-dd}. This requires overall time $O(r \log r)$.
Note that an insert operation may increase by a constant the number of border vertices in at most two clusters and adds an edge to one cluster: since the partition into clusters is recomputed every $\sqrt{r}$ operations, this will guarantee that throughout the sequence of operations each cluster will always have at most $O(r)$ edges and $O(\sqrt{r})$ border vertices. To delete edge $(x,y)$, we remove this edge from the only cluster containing it, and recompute the dense distance graph of this cluster. Once again, this can be carried out in time $O(r \log r)$. Amortizing the initialization over $\sqrt{r}$ operations yields $O(\frac{n \log r + \frac{n}{\sqrt{r}} \log n}{\sqrt{r}} + r \log r)$ time per update.

In order to answer a {\em shortest\_path}$(x,y)$ query, we construct the graph $\overline{G}^{s,t}$ (as defined in Section~\ref{section-cluster}). Note that the distance from $s$ to $t$ in $G_D$ and $\overline{G}^{s,t}$ are equal. Hence, by Corollary~\ref{cor-fast-shortest-paths} the shortest path from $s$ to $t$ can be computed in $O((r+ \frac{n}{\sqrt{r}})\log^2 n)$ time. 

In order to minimize the update time we set $r = n^{2/3}$ and obtain the following theorem. 

\begin{lemma}
\label{lemma:dynamic}
Given a planar graph $G$ with positive edge weights, we can insert edges (allowing changes of the embedding), delete edges and report shortest paths between any pair of vertices in $O(n^{2/3}\log^2 n)$ amortized time per operation.~\footnote{The same worst-case time bounds can be obtained using a global rebuilding technique.\label{footnote-worst-case}}
\end{lemma}

We recall that we can check whether a new edge insertion violates planarity within the bounds of Lemma~\ref{lemma:dynamic}: indeed the algorithm of Eppstein {\em et al.}~\cite{EGIS96} is able to maintain a planar graph subject to edge insertions and deletions that preserve planarity, and allow to test whether a new edge would violate planarity in time $O(n^{1/2})$ per operation. Finally, we observe that not only we have improved slightly the running time over the $O(n^{2/3} \log^{7/3} n)$ time algorithm by Fakcharoenphol and Rao~\cite{fr06}, but our algorithm is also more general. Indeed, the algorithm in~\cite{fr06} can only handle edge cost updates and it is not allowed to change the embedding of the graph. On the contrary, our algorithm can handle the full repertoire of updates (i.e., edge insertions and deletions) and it allows the embedding of the graph to change throughout the sequence of updates.

\medskip

We now turn our attention to dynamic max-flow problems. In particular, given a planar graph $G=(V,E)$ we wish to perform an intermixed sequence of the following operations:

\begin{mylist}{\em shortest\_path$(x,y)$  :}
\litem{\em insert$(x,y,c)$}  add to $G$ an edge of capacity $c$ between vertex $x$ and vertex $y$, provided that the embedding of $G$ does not change;
\litem{\em delete$(x,y)$}  delete from $G$ the edge between vertex $x$ and vertex $y$;
\litem{\em max\_flow$(s,t)$}  return the value of the maximum flow from vertex $s$ to vertex $t$ in $G$.
\end{mylist}

Note that {\em insert} operations are now more restricted than before, as they are not allowed to change the embedding of the graph.
To answer {\em max\_flow} queries in the primal graph $G$ we need to maintain dynamically information about the distances in the
dual graph $G_D$, with the bounds reported in Lemma~\ref{lemma:dynamic}. Unfortunately, things are  more involved as a single edge change in the primal graph $G$ may cause more complicated changes in the dual graph $G_D$. In particular, inserting a new edge into the  primal
graph $G$ results in splitting into two a vertex in the dual graph $G_D$, whereas deleting an edge in the primal graph $G$ implies joining two vertices of $G_D$ into one. However, as edges are inserted into or deleted from the primal graph, vertices in the dual graph
are split or joined according to the embedding of their edges. To handle efficiently vertex splits and joins in the dual graph, we do the following. Let $f$ be a vertex of degree $d$ in the dual graph: we maintain vertex $f$ as a
cycle $C_f$ of $d$ edges, each of cost $0$. The actual edges originally incident to $f$,
are made incident to one of the vertices in $C_f$ in the exact order given by the
embedding. It is now easy to see that in order to join two vertices $f_1$ and $f_2$, we
need to cut their cycles $C_{f_1}$ and $C_{f_2}$, and join them accordingly. This
can be implemented in a constant number of edge insertions and deletions. Similarly, we can support vertex splitting with a constant number of edge insertions and deletions. Additionally, for each cluster $P$, for each pair $h,h'$ of holes we need to compute the dense distance graphs of $P_{\pi_{h,h'}}$ and the minimum-cuts between $b_h$ and $b_{h'}$. However, following Corollary~\ref{corollary-preprocessing}, this does not 
increase the running time of our dynamic algorithm. Note that this information is enough to construct the graphs $\overline{G}^{s,t}$ and $\overline{G}^{s,t}_{\pi}$ in $O(r+\frac{n}{\sqrt{r}})$ time. Moreover, by Theorem~\ref{theorem-holeless-min-cuts} and Theorem~\ref{theorem-general-min-cuts} the min-cut algorithm can be executed on these graphs in $O((r+\frac{n}{\sqrt{r}})\log^3 n)$ time. Setting $r=n^{2/3}$ this yields immediately the main result of this section. 

\begin{lemma}
\label{lemma:dynamicflow}
Given a planar graph $G$ with positive capacities, each operation {\em insert}, {\em delete} and {\em max\_flow} can be implemented in $O(n^{2/3} \log^3 n)$ amortized time.~\footnote{Again, the same worst-case time bounds can be obtained using a global rebuilding technique.}
\end{lemma}

\clearpage
 \bibliographystyle{plain}
 \bibliography{all-gh}

\clearpage

\appendix
\section{Computation of $r$-Partition: Proof of Lemma~\ref{lemma-computing-partition}}
\label{app-clusters}

In order to prove Lemma~\ref{lemma-computing-partition} we will combine Frederickson's~\cite{federickson-87} technique with the following
implication of Section 3.1 and Section~5.1 from~\cite{fr06}. 

\begin{corollary}
\label{corollary-slow-partition}
An $r$-partition of an $n$-node planar graph that already contains $O(\frac{n}{\sqrt{r}})$ border vertices can be computed in $O(n \log n)$ time.
\end{corollary}
\begin{proof}
Actually, Fakcharoenphol and Rao~\cite{fr06} construct in the above time bound a recursive decomposition and in order to get an $r$-partition we
only need to run their algorithm until the size of the clusters drops below $O(r)$. 

Moreover, the recursive decomposition of Fakcharoenphol and Rao~\cite{fr06} assumes that border vertices are present in the decomposed graphs or clusters. Moreover, each time a cluster is split into smaller clusters, the border vertices are split into asymptotically equal parts. The $O(\frac{n}{\sqrt{r}})$ border vertices we start with will be distributed equally into $O(\frac{n}{r})$ clusters. Hence, each cluster in the obtained partition will have additional $O(\frac{n}{\sqrt{r}} \times \frac{r}{n}) = O(\sqrt{r}\,)$ border vertices.
\end{proof}

Now in order to the find the $r$-partition quickly we use the following algorithm:
\begin{itemize}
\item Generate a spanning tree $T$ of the graph;
\item Find connected subtrees of $T$ containing $O(\sqrt{r}\,)$ vertices using a procedure from~\cite{federickson-87};
\item Contract the graph on these subsets, to obtain a simple planar graph $G_s$ with $O(\frac{n}{\sqrt{r}})$ vertices;
\item Using Corollary~\ref{corollary-slow-partition}, find an $r$-division in $G_s$ with $O(\frac{n}{r^{3/2}})$ clusters of size $O(r)$;
\item Expand $G_s$ back to $G$. In $G$ there are $O(n/r)$ clusters $\mathcal{P}_1$ of size $O(\sqrt{r}\,)$ resulting from boundary vertices 
in $G_s$, and  $O(\frac{n}{r^{3/2}})$ clusters $\mathcal{P}_2$ of size $O(r^{3/2})$ resulting from the interior vertices of $G_s$;
\item Apply Corollary~\ref{corollary-slow-partition} to find an $r$-division for clusters in $\mathcal{P}_2$ taking into account the 
border vertices already present in the clusters.
\end{itemize}
It is easy to see that the above procedure requires $O(n \log r + \frac{n}{\sqrt{r}}\log n)$ time. We only need to show that the result is 
a valid $r$-division. The clusters $\mathcal{P}_1$ correspond exactly to the connected subtrees of $T$ of size $O(\sqrt{r}\,)$. They cannot have more then $O(\sqrt{r}\,)$ border vertices and all the border vertices lie on the external face, i.e., they contain one hole. Hence, clusters in $\mathcal{P}_1$ satisfy the properties of an $r$-partition.

\begin{figure}[htp]
\centerline{\psfig{file=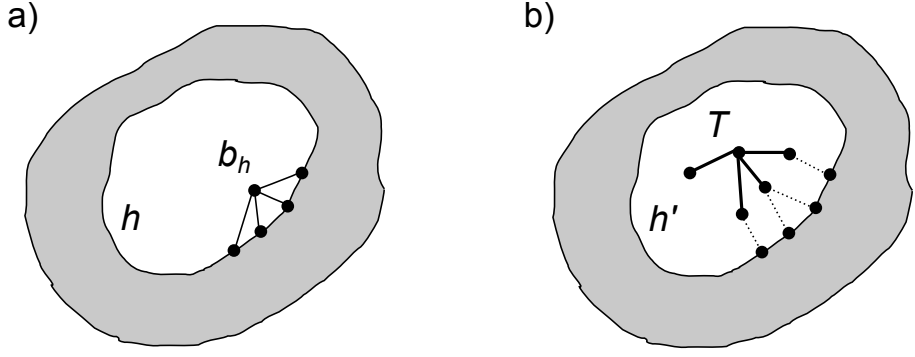,width=12cm}}
\caption{a) Before expanding the border vertex $b_h$ in hole $h$; b) after expanding $b_h$ to $T$ no new hole is created.}
\label{fig:clussterapp}
\end{figure}

Consider the clusters in $\mathcal{P}_2$ before we have applied Corollary~\ref{corollary-slow-partition}. Each cluster is obtained by expanding a cluster in $G_s$ that had $\sqrt{r}$ border vertices. Consider the process of expanding to a subtree $T$ a border vertex $b_h$ lying on a hole $h$ in a cluster $P$. Let $h'$ be the hole obtained from $h$ by removing $b_h$ from $P$. The edges of $T$ were not present in the cluster so the process of expanding $T$ can be seen as gluing $T$ at the side of $h'$. Note that by the connectivity of $T$ no new hole is created and only the hole $h'$ becomes ``smaller''. Moreover, not all of the vertices of $T$ become border vertices of the expanded piece, as some of them do not lie on the side of the new hole (see Figure~\ref{fig:clussterapp}). Nevertheless as $T$ contains $O(\sqrt{r}\,)$ vertices at most $O(\sqrt{r}\,)$ vertices can become border vertices. Hence, in total the pieces in $\mathcal{P}_2$ will have $O(r)$ border vertices. This number satisfies the assumption of Corollary~\ref{corollary-slow-partition}, so pieces obtained by using it satisfy the assumption of $r$-partition. This completes the proof of Lemma~\ref{lemma-computing-partition}.

\section{Computation of Dense Distance Graphs: Proof of Lemma~\ref{lemma-computing-dd}}
\label{appendix-computing-dd}
\begin{proof}
In order to compute dense distance graphs we use the following result by Klein~\cite{klein-05}.

\begin{theorem}[Klein~\cite{klein-05}]
\label{theorem-klein}
Given an $n$-node planar graph with non-negative edge lengths, it takes $O(n \log n)$ time to construct a data structure that supports queries of the following form in $O(\log n)$ time: given a destination vertex $t$ on the boundary of the external face, and given a start vertex $s$ anywhere, find the distance from $s$ to $t$.
\end{theorem}

A cluster $C$ in the $r$-partition has $r$ vertices, $O(\sqrt{r}\,)$ border vertices and a constant number of holes. Border vertices in $C$ lie on one of the holes. In order to compute distances from all border vertices to a given hole $H$ we will apply Theorem~\ref{theorem-klein}. We simply find an embedding of the graph such that $H$ becomes the external face, and query the distances from all border vertices to $H$. There are a constant number of holes in each cluster, and so Klein's data structure will be used a constant number of times for each cluster. On the other hand, there are $O(r)$ pairs of border vertices in each cluster, so we will ask $O(r)$ queries. The time needed to process each cluster is hence $O(r \log r)$, which gives $\frac{n}{r}\times O(r \log r) = O(n \log r)$ time in total.
\end{proof}

\section{Size of Contracted Graphs: Proof of Lemma~\ref{lemma-size-of-graphs}}
\label{appendix-size-of-graphs}
\begin{figure}[thp]
\centerline{
\begin{tabular}{c}
\psfig{file=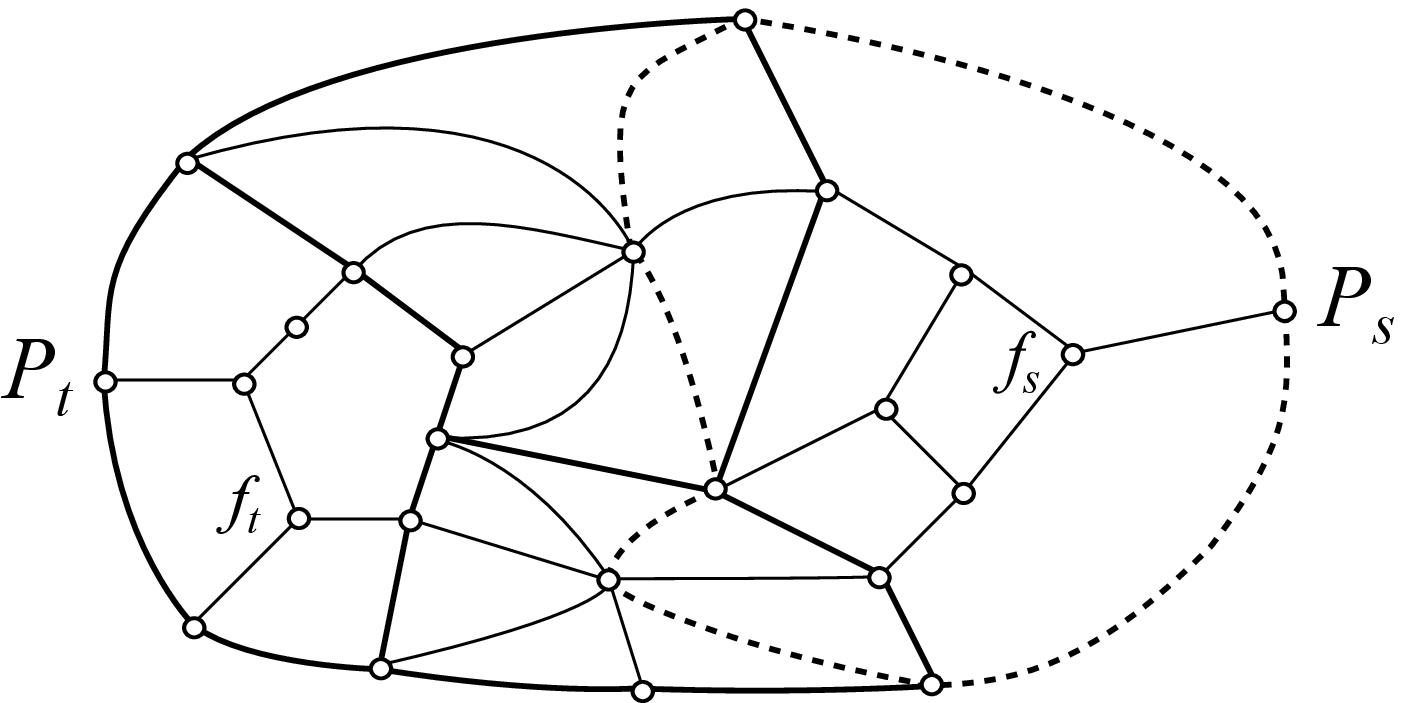,height=4cm}\\(a)\\
\end{tabular}
}
\vspace{7mm}
\centerline{
\begin{tabular}{c}
\psfig{file=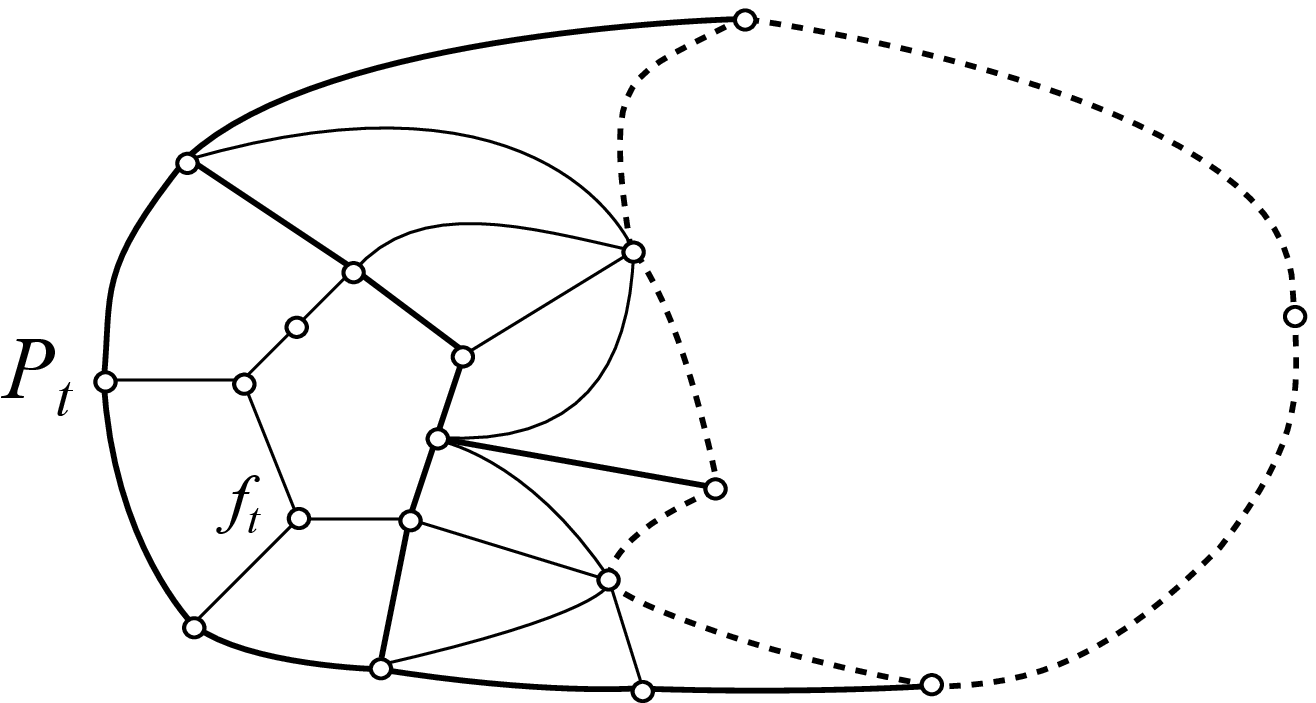,height=4cm}\\(b)\\
\end{tabular}
\begin{tabular}{c}
\psfig{file=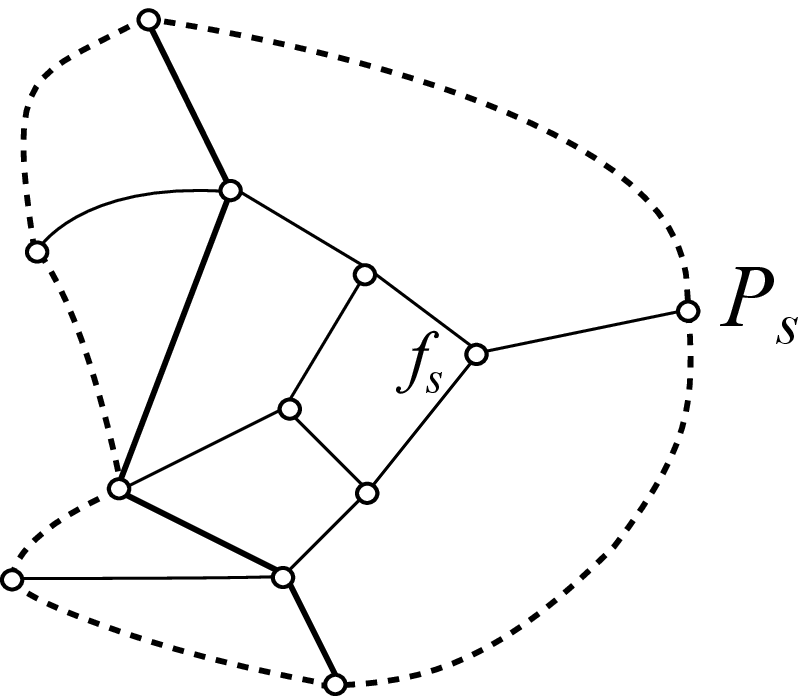,height=4cm}\\(c)\\
\end{tabular}
}
\caption{
(a) The graph $\overline{G}^{s,t}_c$ and the cycle $C_i$. Cluster boundaries are shown in bold, and the  cycle $C_i$ is shown with dashed edges.
(b) The graph $\overline{G}^{i>}_c = \overline{G}^{s,t}_c \cap \extc(C_i)$.
(c) The graph $\overline{G}^{<i}_c = \overline{G}^{s,t}_c \cap \intc(C_i)$. Note that the union of 
$\overline{G}^{i>}_c$ and $\overline{G}^{<i}_c$ contains one more face than $\overline{G}^{s,t}_c$.
}
\label{fig:lemma8}
\end{figure}

\begin{proof}
At the top level of the recursion we have one graph $\overline{G}^{s,t}_c$ that has $O(r + \frac{n}{\sqrt{r}})$ vertices and $O(r + \frac{n}{\sqrt{r}})$ faces. Now consider the case
when the graph $\overline{G}^{s,t}$ is split along $C_i$ into $\overline{G}^{<i}$ and $\overline{G}^{i>}$. Observe that the cycle $C_i$ can be traced in $\overline{G}^{s,t}_c$ when we replace each edge in dense distance graphs for a cluster $P$ by a two-edge path going through the center vertex $v_P$.
In such a case we have $\overline{G}^{<i}_c = \overline{G}^{s,t}_c \cap \intc(C_i)$ and $\overline{G}^{i>}_c = \overline{G}^{s,t}_c \cap \extc(C_i)$. In other words
each time we recurse we split the graph $\overline{G}^{s,t}_c$ along some cycle and then remove degree-two vertices. Note that after splitting
the graph $\overline{G}^{s,t}_c$ along the cycle $C_i$, the number of faces in the union of the resulting graphs increases by exactly one (see Figure~\ref{fig:lemma8}).
There are at most $O(r + \frac{n}{\sqrt{r}})$ recursive calls, so the union all contracted graphs on a given level in the recursive tree has  $O(r + \frac{n}{\sqrt{r}})$ faces as well. Moreover, the vertices in this union graph have degree at least three, because all degree-two vertices are removed. Let $v$, $e$ and $f$ be the number of vertices, edges and respectively faces in the union graph. Now, by Euler's formula the claim of the lemma follows:
$v = 3v - 2v = 3v - 2(2+e-f) \le 2e - 2(2+e -f) \le 2f - 4 = O(r + \frac{n}{\sqrt{r}})$.
\end{proof}

\section{General Clusters}
\label{appendix-general-clusters}
When holes are present in the $r$-partition the skeleton graph $G_{\mathcal{P}}$ is not connected. Hence, the path connecting $b_s$ and $b_t$ cannot use border vertices only. We need to modify the algorithm to allow such paths. We do this as follows.

\begin{figure}[htp]
\centerline{\psfig{file=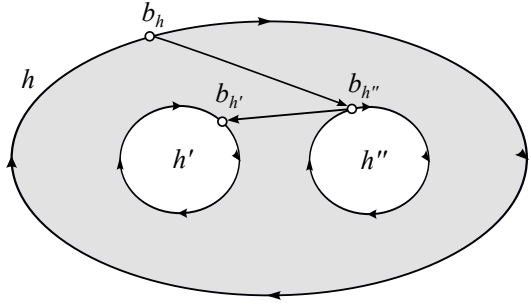,height=4cm}}
\caption{The path $\pi_{h,h'}$.}
\label{fig:holes}
\end{figure}

For each hole $h$ in $P$, we fix a border vertex $b_h$. For each pair of holes $h$, $h'$ in $P$, we fix a path $\pi_{h,h'}$ that starts from $b_h$ and ends in $b_{h'}$, goes through $b_{h''}$ for all holes $h''$ in $P$,
and for all $b_{h''}$ on the path walks around the hole $h''$ passing through all its border vertices (see Figure~\ref{fig:holes}).
These paths are used to do some additional preprocessing for each cluster $P$ in the partition. For each pair of holes $h, h'$ in $P$ we compute dense distance graph for $P_{\pi_{h,h'}}$, and find the minimum cut $C_{h,h'}$ between $b_h$ and $b_{h'}$ in $P$.

\begin{corollary}
\label{corollary-preprocessing}
The additional preprocessing takes $O(n \log r)$ time.
\end{corollary}
\begin{proof}
For each cluster $P$ and each pair of holes the dense distance graph can be computed in the same manner as in Lemma~\ref{lemma-computing-dd}.
On the other hand, minimum-cuts can be found in $O(r\log r)$ time using the algorithm  by Henzinger {\it et al.}~\cite{henzinger-et-al-97}. Hence, over all clusters we need a total of $O(n \log r)$ time.
\end{proof}

\begin{figure}[htp]
\centerline{\psfig{file=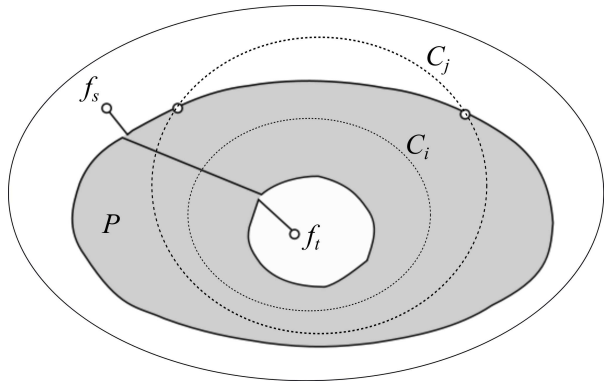,height=5cm}}
\caption{On illustrating Corollary~\ref{corollary-partial-pi}.
A  cut can be either fully contained in one of the pieces $\mathcal{P}_{\pi}$ (such as cut $C_i$) or it must contain one of the border vertices in $\partial \pi$ (such as cut $C_j$).}
\label{fig:corollary6}
\end{figure}

Now in order to connect $f_s$ and $f_t$ we will use paths $\pi_{h,h'}$ whenever we need to pass between two different holes $h,h'$ in a piece $P$.
Let $\mathcal{P}_{\pi}$ be the set of all such pieces on $\pi$. The resulting path is no longer simple, but it will be non-crossing. As shown in Section~\ref{section-non-simple-paths}, our min-cut algorithm can be executed on non-crossing paths as well. We can make the following observation (see Figure~\ref{fig:corollary6}).

\begin{corollary}
\label{corollary-partial-pi}
A minimum $s$-$t$ cut $C$ either contains a vertex in $\partial \pi$ or is fully contained in one of the pieces $\mathcal{P}_{\pi}$.
\end{corollary}
\begin{proof}
If the cycle $C$ contains a vertex in $\partial \pi$, we are done. Assume that it does not contain any vertex from $\partial \pi$. In order to be a cut, it has to cross path $\pi$ and it can do it so in one of $\pi_{h,h'}$. Then it has to be fully contained in the corresponding $P$ as by the construction of $\pi_{h,h'}$ or border vertices of $P$ lie on $\pi$.
\end{proof}

Using  Corollary~\ref{corollary-partial-pi}, we can find the minimum cut in two phases.
First, let $C_{i}$ be the smallest of the cuts $C_{h,h'}$ in $\mathcal{P}_{\pi}$ for $b_h, b_h' \in \pi$. Second, run the algorithm from previous section on a path $\partial \pi$ in $\overline{G}^{s,t}$ to find a cut $C_b$. Finally, return the smallest of the cuts $C_i$ and $C_b$.
The running time of the above algorithm is the same as in Theorem~\ref{theorem-holeless-min-cuts}.

\begin{theorem}
\label{theorem-general-min-cuts}
Let $G$ be a flow network with source $s$ and sink $t$. The minimum cut between $s$ and $t$ in $G$ can be computed in $O(n \log r + (r+\frac{n}{\sqrt{r} })\log^3 n)$ time. By setting $r = \log^8 n$, we obtain an $O(n \log \log n)$ time algorithm.
\end{theorem}

\end{document}